# Projected Effects of Naloxone and Buprenorphine Scale-Up on Opioid Overdose Mortality in Six Pennsylvania Counties

Abdulrahman A. Ahmed[1], M. Amin Rahimian[*,1], Qiushi Chen[2], Praveen Kumar[*,3]

[1]Department of Industrial Engineering, University of Pittsburgh, Pittsburgh, USA

[2]Harold and Inge Marcus Department of Industrial and Manufacturing Engineering, The Pennsylvania State University, University Park, USA

[3]Department of Health Policy and Management, School of Public Health, University of Pittsburgh, Pittsburgh, USA

## Abstract

**Background:** Opioid overdose deaths remain a severe public health crisis in the US, with heterogenous burden across counties that differ in epidemic trajectory, baseline resources, and local context. While harm reduction through naloxone distribution and buprenorphine treatment are both evidence-based strategies, limited information on county-level effects hinders the ability of policymakers to prioritize resources across counties.

**Methods:** We developed a simulation model of opioid use disorder (OUD), calibrated separately to six Pennsylvania counties spanning large urban (Allegheny, Philadelphia), intermediate-sized (Erie, Dauphin), and rural (Clearfield, Columbia) settings. We projected county-specific overdose mortality trajectories under three levels of increase in buprenorphine dispensing and naloxone distribution (10%, 20%, and 30% above each county's baseline), over a five-year horizon from 2025 to 2029.

**Results:** A 30% increase in naloxone distribution above projected county baseline levels was projected to reduce 2029 overdose deaths by approximately 70% (95% uncertainty interval, UI: 55-81%) in Allegheny County, 11% (95% UI:4-17%) in Erie County, and 28% (95% UI:2-64%) in Clearfield County. The relative projected benefits of naloxone and buprenorphine varied by county; buprenorphine produced larger reductions in Erie and Dauphin.

**Conclusions:** The same proportional increase in naloxone distribution produced substantially different projected mortality reductions across the six Pennsylvania counties. County-level context is important for informing harm reduction and treatment prioritization at the county level.

*Correspondence to: M. Amin Rahimian (rahimian@pitt.edu) and Praveen Kumar (prk52@pitt.edu).

## 1. Introduction

The United States opioid epidemic has claimed more than 750,000 lives over the past two decades, evolving through waves of prescription opioids, heroin, and now illicitly manufactured fentanyl and its analogues (Ciccarone, 2019; Garnett et al., 2024; Volkow & Blanco, 2021). The epidemic has been geographically heterogeneous: in 2023, drug overdose mortality ranged from 9.0 per 100,000 population in Nebraska to 81.9 per 100,000 in West Virginia, and the geographic and demographic distribution of overdose deaths has shifted substantially across epidemic waves (Centers for Disease Control and Prevention, 2025).

A wide range of strategies has been deployed in response to the epidemic, including prescription drug monitoring programs, linkage-to-care programs initiated in emergency departments, syringe service programs, and expanded access to medications for opioid use disorder (MOUD) (D'Onofrio et al., 2015; Pitt et al., 2018; Wakeman et al., 2020). One of the promising strategies is the distribution of naloxone, an opioid antagonist distributed as a harm reduction strategy that reverses overdose when administered promptly, with population-level studies demonstrating significant reductions in overdose mortality at scale (Coffin & Sullivan, 2013; Walley et al., 2013). Another promising strategy is the use of buprenorphine, a partial opioid agonist used as a MOUD, which reduces illicit opioid use, overdose risk, and all-cause mortality through sustained treatment engagement (Mattick et al., 2014; Sordo et al., 2017). Both strategies have been adopted nationally, supported by legislative changes and expanded public health programming. Yet for a county health department, the relevant question is not whether naloxone or buprenorphine work, as their clinical effectiveness has been well established, but how much overdose mortality reduction a specific proportional increase in local dispensing is expected to produce at the population level.

This question is challenging because counties differ not only in how severely the opioid epidemic has affected them but also in the resources available for existing interventions. Understanding these context-specific differences in intervention responsiveness is central to what has been described as precision public health: moving from population-average recommendations toward interventions tailored to local capacity and need (Khoury & Evans, 2015).

Several models have been developed to project opioid intervention effects, but differ substantially in geographic scope, modeling approach, and how intervention effects are estimated. At the national level, the SOURCE model (Lim et al., 2022; Stringfellow et al., 2022) identified naloxone and buprenorphine among the highest-impact strategies at the national level, but estimates a single naloxone efficiency parameter applied uniformly to the entire US population, and therefore does not explicitly estimate county-specific effects. Linas et al. (2021) applied a decision-analytic model at the state level in Massachusetts, with naloxone effectiveness represented as a literature-derived survival multiplier. Chhatwal et al. (2023) fitted a state-transition model to four states, Kentucky, Massachusetts, New York, and Ohio, but represented naloxone as a uniform 10% reduction in overdose mortality rates applied identically across all settings, a value chosen from prior literature, precluding any mechanism by which local distribution history could produce different effects. Irvine et al. (2022) developed a mathematical model to estimate naloxone need across 12 US states, but grouped states by dominant opioid epidemic type, fentanyl, heroin, and prescription opioid, sharing parameters across jurisdictions within each group rather than fitting to individual states, and modeled naloxone as absolute kit counts through a saturating function rather than as a county-level dispensing rate linked to local mortality targets. Cerdá et al. (2024) used agent-based models to examine combinations of naloxone distribution and buprenorphine initiation and retention needed to achieve a 40% reduction in overdose deaths in eight New York counties. We extend this line of research by estimating the effects of increasing buprenorphine and naloxone in six Pennsylvania counties over 2025-2029.

In this study, we developed a simulation model of opioid use disorder (OUD) adapted to six Pennsylvania counties spanning large urban, intermediate-sized, and rural settings. We projected county-specific overdose mortality trajectories under realistic naloxone and buprenorphine intervention scenarios and quantified how identical proportional increases in these interventions translate to substantially different mortality reductions across counties.

## 2. Methods

### 2.1 Overview

We developed a Markov cohort model (Siebert et al., 2012) that represents progression through opioid use, treatment, and overdose death, with model inputs informed by published clinical and epidemiological studies and county-level data from various sources. We calibrated the model to six Pennsylvania counties separately, selected to reflect heterogeneity in population size and urban-rural composition (**eTable 1**). We projected opioid overdose deaths over a five-year horizon (2025-2029) under different intervention levels.

### 2.2 OUD Model

The simulation model is based on the Markov framework with seven states that track the population through different health states over time to estimate the effects of harm reduction and treatment interventions on opioid overdose mortality. The model captures the core pathway through which individuals enter and progress through opioid use: from non-use through prescription opioid (PO) use and misuse, into opioid use disorder (OUD), and onward to treatment (MOUD), or overdose death (**Figure 1**). This structure was chosen to represent the transitions most relevant to naloxone and buprenorphine interventions: naloxone acts on the overdose death transition, and buprenorphine acts on the treatment entry transition, allowing the model to isolate each intervention's mechanism. The model updates each month (i.e., cycle length of 1 month), allowing it to capture the short-term dynamics of changing overdose risk affected by interventions. The initial distribution of population across health states was based on prevalence estimates from the National Survey on Drug Use and Health (NSDUH) and consultation with public health experts (**eTable 2**).

## 2.3 Parameter Estimation

We parameterized the model for six Pennsylvania counties spanning large urban (Allegheny, Philadelphia), intermediate-sized (Dauphin, Erie), and rural (Clearfield, Columbia) contexts. For most transition probabilities between health states, values were estimated from published studies and shared across counties (**Table 1**). However, the following three transition probabilities – from non-user to PO use ($p_1$), from OUD to MOUD ($p_2$), and from OUD to overdose death ($p_3$) – depended on county specific availability of opioid prescriptions, buprenorphine, and naloxone. These three probabilities were estimated by calibrating the model to county specific overdose death rates, as described in the '*Model Calibration*' section.

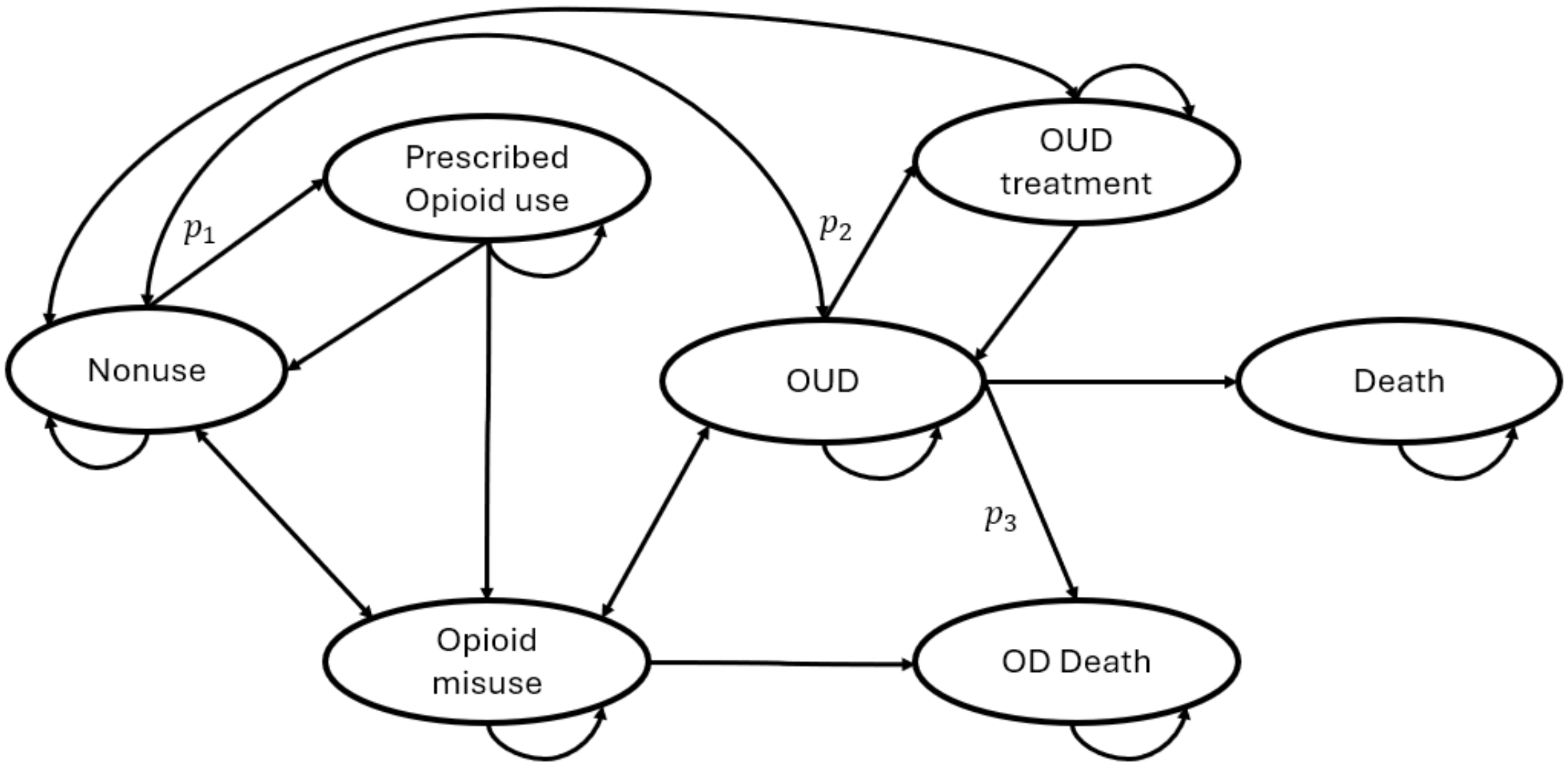

*Figure 1. State transition diagram for the seven-state OUD Markov model. NU = non-use; PO = prescription opioid use; MU = misuse; OUD = opioid use disorder; MOUD = medications for OUD. Overdose death and death from other causes are absorbing states.*

**Table 1. Transition Probability Parameters for the OUD Model**

| Parameter | Value | Source |
|---|---|---|
| *Transitions from Nonuser State* | | |
| Non-use to PO use ($p_1$)[1, 2] | 0.0003-0.4479[3] | Calculated (eTable 4) |
| Non-use to Misuse | 0.001 | (Compton et al., 2013) |
| Non-use to OUD | 0.0008 | (Compton et al., 2013) |
| *Transitions from Prescription Opioid (PO) Use State* | | |
| PO use to Nonuse | 0.815 | (Centers for Disease Control and Prevention, 2018) |
| PO use to Misuse | 0.05 | (Manchikanti et al., 2006) |
| PO use to OUD | 0.0327 | (Fishbain et al., 2008) |
| *Transitions from Misuse State* | | |
| Misuse to Nonuse | 0.039 | (Compton et al., 2013) |
| Misuse to OUD | 0.023 | (Compton et al., 2013) |
| Misuse to Overdose Death | $1.31 \times 10^{-7}$ | Assumption |
| *Transitions from OUD State* | | |
| OUD to Nonuse | 0.0162 | (Compton et al., 2013) |
| OUD to Misuse | 0.013 | (Compton et al., 2013) |
| OUD to MOUD ($p_2$) [1, 2] | 0.000-0.122[3] | Calculated (eTable 4) |
| OUD to Overdose Death ($p_3$) [1, 2] | 0.0003-0.00498[3] | Calculated (eTable 4) |
| OUD to Death, non-overdose | 0.002818 | (Kochanek et al., 2019) |
| *Transitions from MOUD State* | | |
| MOUD to Nonuse | 0.009 | (Weiss et al., 2015) |
| MOUD to OUD | 0.0089 | (Weiss et al., 2015) |
| *General Mortality* | | |
| All-cause mortality (monthly, ages 20-60) | 0.000238 | (Kochanek et al., 2019) |
| *Covariates range across counties lowest – highest* | | |
| Naloxone dispensing rate[4] | 39-5989[3] | IQVIA |
| Buprenorphine dispensing rate[4] | 1548-25286[3] | IQVIA |
| Opioid dispensing rate[4] | 38357 - 68955[3] | IQVIA |
| % of seizures involving fentanyl | 0.001-0.26[3] | NFLIS (state-level) |

**Note.** 1) PO = prescription opioid; OUD = opioid use disorder; MOUD = medications for OUD; NFLIS = National Forensic Laboratory Information System

2) $p_1$, $p_2$, and $p_3$ are covariate-dependent and estimated via county-specific calibration (**eAppendix 2**).

3) The range represents minimum and maximum probability across 6 counties

4) The dispensing rates are expressed per 100,000 population

### 2.3.1 Model Calibration

The county-specific effects of harm reduction and treatment receipt on overdose mortality were estimated by calibrating the model to annual county-level overdose death rates for 2018-2022 from CDC WONDER (CDC WONDER, 2024), specifically three transition probabilities: from non-user to PO user ($p_1$), from OUD to MOUD ($p_2$), and from OUD to overdose death ($p_3$). Each of these transition probabilities was parameterized as a logistic function dependent on county-level covariates such as opioid, buprenorphine, and naloxone dispensing rates obtained from IQVIA, and state-level fentanyl seizure data obtained from the National Forensic Laboratory Information System (**eAppendix 2**).

We employed the Incremental Mixture Importance Sampling (IMIS) algorithm (Raftery & Bao, 2010), a Bayesian calibration approach (Menzies et al., 2017), to obtain posterior samples of the calibrated parameters (**eTable 4).** We specified weakly informative uniform priors with bounds tailored to each county (**eTable 3**). The overdose death rates for 2023-2024 were reserved for validation. Further details on calibration are available in eAppendix 2.

### 2.4 Intervention Scenarios

We evaluated two evidence-based interventions: i) harm reduction through increased naloxone distribution, and ii) treatment receipt through increased buprenorphine dispensing, as proportional increases of 10%, 20%, and 30% above each county's baseline dispensing level. Proportional increases were chosen as the policy-relevant metric because they represent feasible scale-up from each county's existing distribution infrastructure, for example, a 10% increase in dispensing from existing program levels. Per-capita absolute targets were not used because they would implicitly require counties with limited existing infrastructure to achieve the same absolute volume as counties with established programs, ignoring baseline differences in distribution capacity. For the 2025-2029 projection period, baseline covariate values (buprenorphine and naloxone dispensing) were projected by fitting a linear trend to county-specific observed values from 2018-2024 (**eFigure 2**); intervention scenarios were then applied as proportional increases relative to each county's projected baseline levels.

### 2.5 Model Outcomes

The primary outcome was annual overdose death rates per 100,000 population projected for 2025-2029 under each intervention scenario. Overdose deaths averted relative to the no-increase baseline were computed sample-wise across 250 posterior draws and reported as mean reductions with 95% uncertainty intervals. All outcomes are reported per 100,000 population.

## 3. Results

The model was calibrated separately to each county using annual overdose death rates from 2018 to 2022 as calibration targets, with 2023-2024 serving as an out-of-sample validation period for all counties, except for Columbia and Clearfield in 2024 where data were suppressed due to fewer than 10 overdose deaths observed that year. Calibration and validation results are shown in **eFigure 1**. Across all six counties, the model closely reproduced observed overdose death rates across most years and counties, including validation years, indicating that it adequately captured the underlying trends. The exception was Allegheny County, where the model underestimated observed overdose deaths in the validation years (2023 and 2024). Therefore, we conducted a sensitivity analysis in which the Allegheny County model was calibrated using data from 2018–2024 to estimate reductions in overdose death rates under the intervention scenarios (**eAppendix 3, eFigure 3**).

### 3.1 Overdose death rates in the baseline scenario

Under the no-intervention baseline, projected overdose mortality is expected to decline in most counties over the 2025-2029 horizon if recent dispensing trends continue (**Figure 2**). The steepest projected declines were in Allegheny and Dauphin, where baseline 2029 rates decreased to approximately 2 [95% UI: 0.3-6] and 3 [95% UI: 1.5-7.1] per 100,000, respectively (in the complementary calibration that additionally included 2023 and 2024 as calibration targets, the Allegheny baseline 2029 rate was 5, 95% UI: 2.9-8.3, **eAppendix 3**). Clearfield, Columbia and Erie showed more moderate declines to approximately 8 [95% UI: 3.9-17.4], 7 [95% UI: 3.2-32.3] and 15 [95% UI: 7.7-33.9] per 100,000 respectively, while Philadelphia remained higher and roughly stable at approximately 31 (95% UI: 12.9-47.3) per 100,000. These baseline trajectories provide the reference against which the intervention scenarios below are compared.

### 3.2 Scaling up naloxone and buprenorphine separately

**Figures 2** and **3** show projected overdose mortality trajectories under scenarios with increasing naloxone and buprenorphine, respectively, demonstrating the distinct mechanisms and magnitudes of each intervention type.

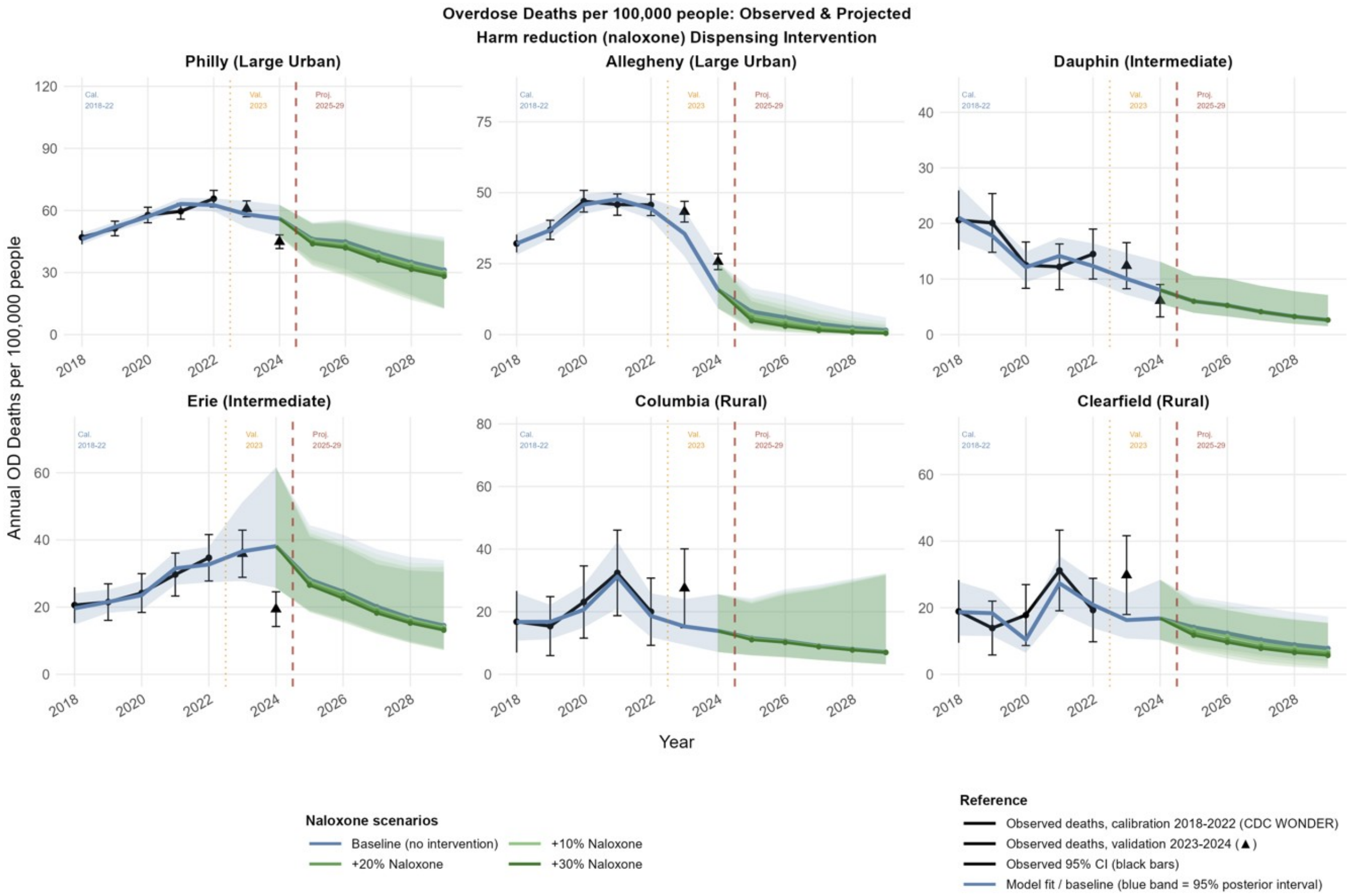

*Figure 2. Projected overdose death rates per 100,000 (2025-2029) under naloxone distribution interventions (10%, 20%, and 30% increases above projected county baseline) by county. Black solid line: observed 2018-2024 except for Columbia and Clearfield till 2023 (CDC WONDER). Blue solid line with shading: model predicted overdose death rates in the baseline scenario (median, 95% uncertainty interval). Triangle dots: model validation 2023-2024. Colored lines: intervention scenarios (light to dark green: +10% to +30% naloxone). Red vertical dashed line: start of projection period (2025-2029).*

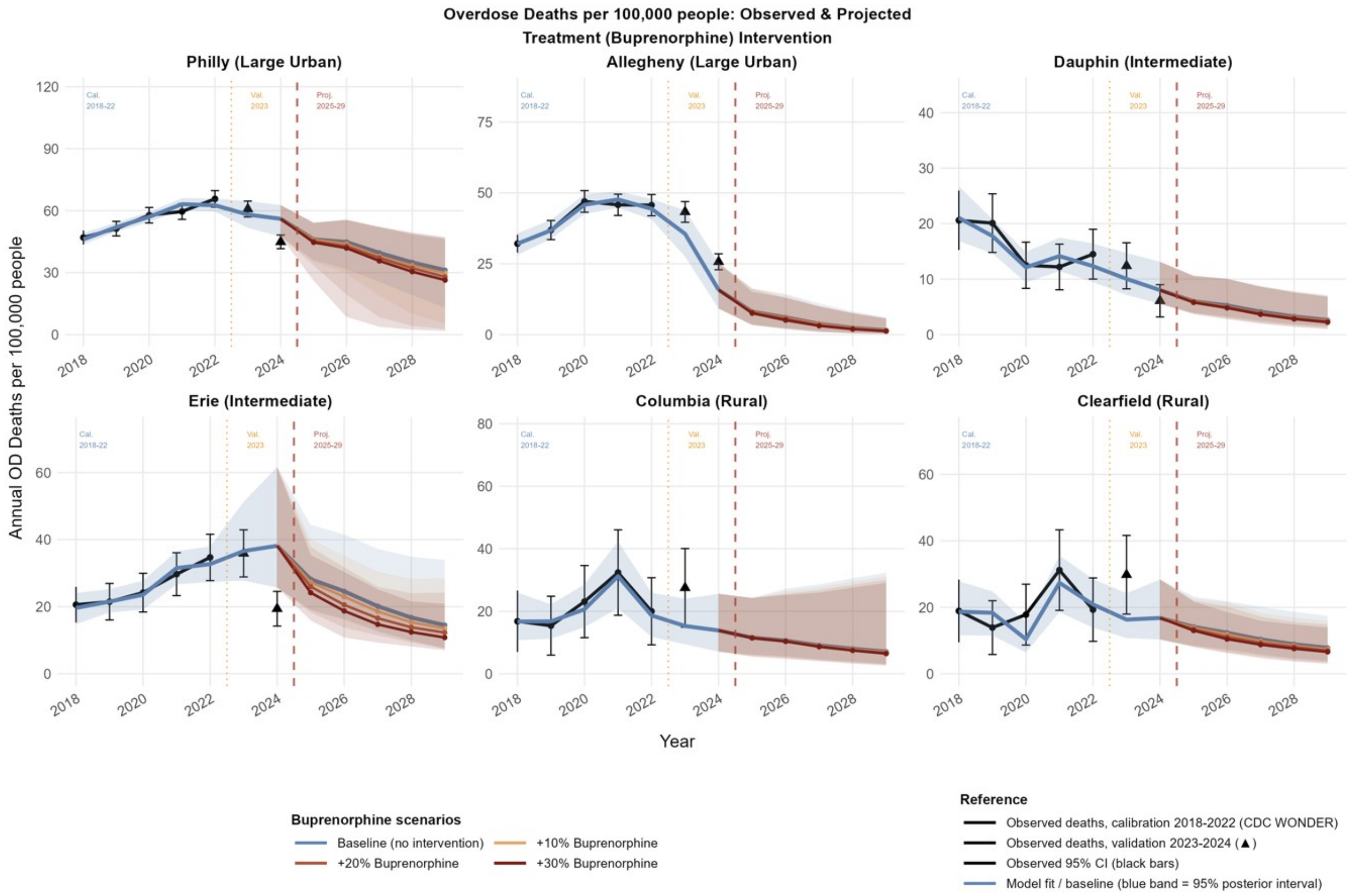

*Figure 3. Projected overdose death rates per 100,000 (2025-2029) under buprenorphine interventions (10%, 20%, and 30% increases above projected county baseline) by county. Black solid line: observed 2018-2024 except for Columbia and Clearfield till 2023 (CDC WONDER). Blue solid line with shading: model predicted overdose death rates in the baseline scenario (median, 95% uncertainty interval). Triangle dots: model validation 2023-2024. Colored lines: +10% to +30% buprenorphine (orange to dark red). Red vertical dashed line: start of projection period (2025-2029).*

Naloxone scenarios (**Figure 2**) show consistent declines in most counties but the magnitude varied widely. Allegheny showed the strongest response, with the +30% scenario producing an approximately 70% [95% UI: 55-81%] reduction in the 2029 projected death rate relative to the no-intervention baseline. Results for Allegheny were similar (59% [95% UI: 51-65%]) in the sensitivity analysis using the 2018–2024 calibration period (**eFigure 4)**, supporting the robustness of the estimated intervention effect. Clearfield and Erie showed intermediate responses (approximately 28% [95% UI: 2-64] and 11% [95% UI: 4-17] at +30%, respectively), while Philadelphia and Columbia showed smaller reductions (approximately 10% [95% UI: 1-22] and 2% [95% UI: 0-3]). In Dauphin, naloxone scale-up produced almost no projected change (under 1% at +30%).

Buprenorphine scenarios (Figure 3) produced smaller projected reductions in mortality compared to naloxone in several counties, likely reflecting buprenorphine's indirect pathway to reducing overdose mortality through increased treatment entry ($p_2$), At +30%, projected 2029 reductions were approximately 23% (95% UI: 0-56) in Allegheny, 20% (95% UI: 0-58) in Erie, 15% (95% UI: 0-45) in Dauphin, 15% (95% UI: 1-40) in Clearfield, and 10% (95% UI: 0-91) in Philadelphia. In a sensitivity analysis using 2018-2024 data for model recalibration, the projected reduction in Allegheny was 42% (95% UI: 6-63; **eFigure 4**). In Erie and Philadelphia, the projected buprenorphine response matched or modestly exceeded the naloxone response at the same intervention level, whereas in Columbia buprenorphine scale-up showed the smallest projected change.

### 3.3 Scaling up both naloxone and buprenorphine simultaneously

**Figure 4** shows projected 2025-2029 trajectories under scenarios that scale up both interventions jointly. The added value of buprenorphine over naloxone alone varied across counties. In Allegheny, the combined +30%/+30% intervention increase reached an approximately 80% (95% UI: 59–89) projected reduction by 2029, compared with 70% [95% UI: 55-81%] for naloxone alone and 23% [95% UI: 0–56%] for buprenorphine alone, indicating a largely additive contribution. In Erie, where naloxone alone produced a more limited reduction, the joint scale-up reached approximately 28% (95% UI: 5–64), with buprenorphine contributing the larger share. In counties where a single intervention already produced little change, joint scale-up added correspondingly little.

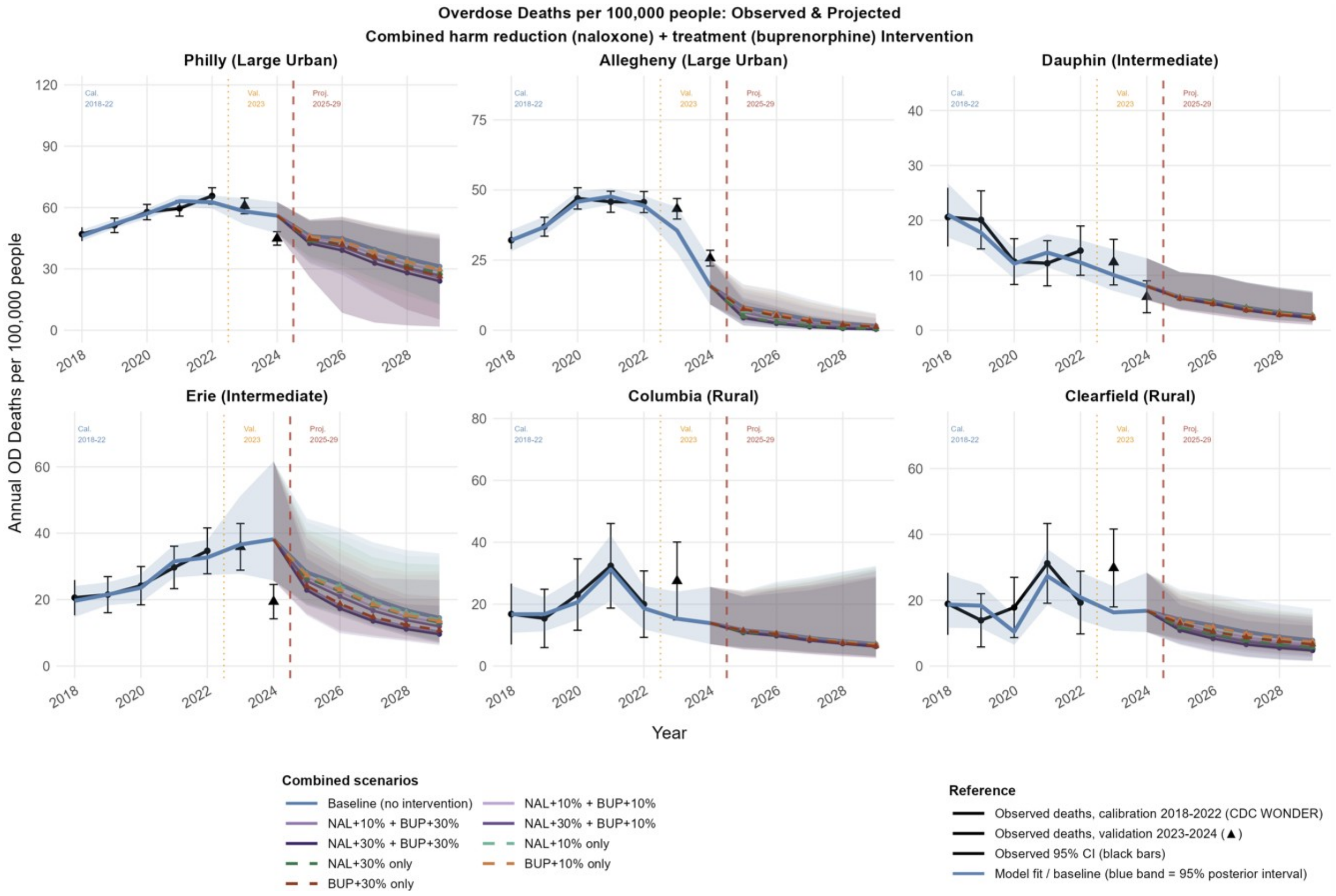

*Figure 4. Historical observed overdose death rates (2018-2022; black solid line with points) and model-projected rates for 2025-2029 under combined harm reduction and treatment scenarios. Colored solid lines: combined scenarios at four levels (NAL+10%/BUP+10%, NAL+10%/BUP+30%, NAL+30%/BUP+10%, NAL+30%/BUP+30%). Dashed colored lines: single-arm reference scenarios (naloxone alone, buprenorphine alone at same percentage levels). Blue solid line with shading: model predicted overdose death rates in the baseline scenario (median, 95% uncertainty interval). Triangle dots: model validation (2023-2024, out-of-sample). Red vertical dashed line: start of projection period (2025-2029).*

## 4. Discussion

We developed a simulation model to estimate county-specific effects of harm reduction (naloxone) and treatment (buprenorphine) interventions on overdose mortality over a 2025-2029 projection horizon in six counties of Pennsylvania. Three key findings emerge. First, both interventions produced meaningful projected reductions in overdose deaths, but which intervention produced the larger reduction was county-specific rather than uniform: harm reduction through naloxone produced the larger effect in some counties (e.g., Allegheny and Clearfield), while increased treatment engagement through buprenorphine matched or exceeded it in others (e.g., Erie and Philadelphia). Second, substantial heterogeneity in naloxone effect sizes was observed across counties: a 30% increase in naloxone distribution was projected to reduce overdose deaths in 2029 by approximately 70% (95% UI: 55-81) in Allegheny County, compared to approximately 11% (95% UI: 4-17) in Erie County. Third, this heterogeneity did not track county characteristics such as urbanicity or population size. This suggests that projected benefit may not be simply inferred from these descriptors alone, and that county-specific assessment may be useful when prioritizing interventions.

These findings are consistent with and extend prior simulation evidence in several respects. Naloxone's substantial projected benefits in several counties are consistent with national modeling evidence identifying naloxone distribution among the highest-impact strategies for reducing overdose mortality (Stringfellow et al., 2022), and with evidence that buprenorphine's mortality benefit is mediated by sustained treatment retention, with mortality risk rising sharply when treatment is discontinued (Sordo et al., 2017). Geographic heterogeneity in intervention effects has been documented in prior work, including Cerda et al. (2024) across eight New York counties and Irvine et al. (2022) across 12 US states by opioid epidemic type. Our findings extend this evidence to six Pennsylvania counties by estimating how incremental 10%, 20%, and 30% increases in naloxone or buprenorphine dispensing translate into projected mortality reductions, with posterior uncertainty intervals.

The county-specific projections produced here offer locally grounded evidence to support resource allocation decisions by county health departments and public health planners. Our results demonstrate that the same proportional increase in naloxone or buprenorphine dispensing yields substantially different projected reductions across counties, meaning that uniform, population-proportional strategies are unlikely to produce uniform outcomes. Where national models can only recommend strategies that work on average, county-level estimates allow decision-makers to identify which intervention is expected to produce the greatest mortality reduction given their county's position in the epidemic trajectory. These estimates are particularly relevant in the context of opioid settlement funds, where states and counties face consequential choices about how to prioritize harm reduction relative to treatment across heterogeneous local contexts (Sharfstein & Olsen, 2020; Skinner et al., 2024), as county-specific estimates of intervention responsiveness can inform these decisions.

### 4.1 Limitations

Several limitations should be acknowledged. First, the model evaluates only naloxone and buprenorphine as two representative interventions commonly considered in opioid policies; other evidence-based strategies including methadone maintenance, naltrexone, syringe service programs, and safe prescribing were not modeled, and the optimal combination of interventions across the full response landscape remains an open question. Second, the model was calibrated to a single outcome, annual county-level overdose death rates, and the lack of other reliable opioid and substance use-related measures at the county-level granularity such as OUD prevalence, treatment admissions, or nonfatal overdose counts limits parameter identifiability. Third, fentanyl data were available only at the state level, potentially obscuring county-specific supply dynamics that drive overdose risk. Fourth, estimates of naloxone and buprenorphine supply are based on dispensing data from the IQVIA Xponent database, which primarily captures retail pharmacy channels and excludes other sources of supply. Thus, our estimates likely underestimate actual availability, particularly for naloxone, for which non-retail channels now account for a substantial share of supply. Because we did not have access to data from these other channels, dispensing data should be interpreted as an indicator of changes in naloxone and buprenorphine supply over time, rather than as a measure of absolute supply. Fifth, the model treats county populations as homogeneous aggregates; models that incorporate demographic stratification by race, ethnicity, or socioeconomic status would better capture within-county disparities in intervention access and overdose risk.

## 5. Conclusion

The opioid epidemic continues to strike counties differently; this study found substantial county-level variation in the impact of harm reduction and buprenorphine treatment on overdose deaths. The relative projected benefits of naloxone and buprenorphine varied across counties, supporting locally informed intervention planning. Lastly, combining these interventions is likely to yield greater benefits than implementing either strategy alone.

## Declarations

**Conflicts of interest:** The authors declare no conflicts of interest.

**Funding:** This work has been funded by the contract 75D30121C12574 from the Centers for Disease Control and Prevention (CDC) and by the National Science Foundation under grant agreement CMMI-2240408. The findings and conclusions in this work are those of the authors and do not necessarily represent the official position of the CDC. Any opinions, findings and conclusions or recommendations expressed in this material are those of the author(s) and do not necessarily reflect the views of the U.S. National Science Foundation.

**Declaration of generative AI and AI-assisted technologies in the manuscript preparation process.** During the preparation of this work the author(s) used Claude and ChatGPT to assist with plots and in refining sentence structure, improving clarity, and ensuring conciseness in the manuscript. After using these tools, the author(s) reviewed and edited the content as needed and take full responsibility for the content of the published article.

**Data availability:** Our analysis relies on multiple data streams that collectively capture OUD dynamics at the county level. Monthly county-level dispensing rates for prescription opioids, naloxone, and buprenorphine were obtained from the IQVIA dataset. These data reflect prescriptions dispensed across retail, mail-order, and long-term care pharmacies. This data is not publicly available. Access can be requested through IQVIA at https://www.iqvia.com/insights/the-iqvia-institute/available-iqvia-data. County-level overdose death counts were collected from the CDC Wide-Ranging Online Data for Epidemiologic Research, identified using International Classification of Diseases, 10th Revision (ICD–10) codes for opioid-related poisoning (X40–X44, X60–X64, X85, Y10–Y14, T40.0–T40.4, T40.6). In addition, fentanyl seizure data were obtained from the National Forensic Laboratory Information System (NFLIS) and can be accessed at https://www.nflis.deadiversion.usdoj.gov.

**Code Availability:** The simulation model was developed in R. The R scripts used to generate the results reported in this study are publicly available at https://github.com/abdulrahmanfci/county-oud-model.

**Consent and Approval Statement**: The Institutional Review Board (IRB) at the University of Pittsburgh determined that this study did not constitute human subjects research.

# Supplementary Materials

## eAppendix 1: County Selection and Model Structure

We selected six Pennsylvania counties to capture heterogeneity in population size, urbanicity, and opioid burden. This strategy enables examination of how intervention effectiveness varies across counties with distinct demographic profiles, baseline dispensing rates, and opioid epidemic dynamics.

**eTable 1. Characteristics of six Pennsylvania Counties Selected for Analysis**

| County | Population | Urbanicity | Opioid overdose deaths | | | | | | |
|---|---|---|---|---|---|---|---|---|---|
| | | | 2018 | 2019 | 2020 | 2021 | 2022 | 2023 | 2024 |
| Allegheny | 1,249,020 | Large urban | 32.1 | 36.9 | 47.0 | 45.8 | 45.7 | 43.3 | 25.7 |
| Philadelphia | 1,583,219 | Large urban | 47.0 | 51.3 | 57.8 | 59.6 | 65.7 | 60.8 | 44.9 |
| Dauphin | 276,645 | Intermediate | 20.6 | 20.1 | 12.5 | 12.2 | 14.5 | 12.4 | 6.1 |
| Erie | 279,609 | Intermediate | 20.6 | 21.5 | 24.2 | 29.7 | 34.7 | 35.9 | 19.4 |
| Columbia | 66,647 | Small/rural | 16.8 | 15.4 | 23.1 | 32.4 | 20.0 | 27.5 | NA |
| Clearfield | 81,708 | Small/rural | 18.9 | 13.9 | 17.8 | 31.2 | 19.3 | 29.8 | NA |

**Note.** Opioid overdose death rates per 100,000 population (CDC WONDER).

NA – Not available

## eTable 2. Initial State Distribution Across Health States at Model Initialization

| Health State | Proportion |
|---|---|
| Nonuser | 0.909 |
| Prescription opioid user | 0.006 |
| Misuser | 0.040 |
| Opioid use disorder (OUD) | 0.040 |
| OUD in treatment | 0.005 |
| Overdose death | 0.000 |
| Death (other causes) | 0.000 |

**Note.** Proportions sum to 1.0.

Informed by the National Survey on Drug Use and Health prevalence estimates and expert consultation.

**eAppendix 2: Bayesian Calibration Details**

The three covariate-dependent transition probabilities were parameterized as logistic functions of standardized county-level covariates:

$$\text{logit}(p_1) = \beta_0 + \beta_1 \cdot \text{opioid dispensing rate}$$

$$\text{logit}(p_2) = \beta_2 + \beta_3 \cdot \text{buprenorphine dispensing rate}$$

$$\text{logit}(p_3) = \beta_4 + \beta_5 \cdot \text{\% of seizures involving fentanyl} - \beta_6 \cdot \text{naloxone dispensing rate}$$

where $p_1$ is the monthly transition probability from non-user to prescription opioid user, $p_2$ is the monthly transition probability from OUD to MOUD, and $p_3$ is the monthly transition probability from OUD to overdose death. Parameters $\beta_0$ through $\beta_6$ are the seven calibrated intercept and slope coefficients. We calibrated these seven parameters using IMIS (Raftery and Bao, 2010), a Bayesian algorithm that iteratively refines the sampling distribution to approximate the posterior efficiently. Calibration targets were annual county-level overdose death rates per 100,000 population from 2018 to 2022, obtained from CDC WONDER. All covariates were standardized (z-scores) using county-specific means and standard deviations computed over the 2018-2022 calibration period. We specified weakly informative uniform priors with bounds tailored to each county (eTable 3).

Opioid, buprenorphine, and naloxone dispensing rates were obtained from the IQVIA Xponent database, which reports the number of prescriptions dispensed through retail pharmacy channels at the county level. Each dispensing rate is expressed as the number of prescriptions dispensed per 100,000 population per year: the opioid dispensing rate for prescription opioids, and the buprenorphine and naloxone dispensing rates for each respective medication. The proportion of seizures involving fentanyl was derived from state-level seizure data from the National Forensic Laboratory Information System (NFLIS) and used as an indicator for the local supply of illicit synthetic opioids; because NFLIS reports at the state level, this measure is common to all counties within a state.

Opioid, buprenorphine, and fentanyl covariates entered the model as their annual values. Naloxone was treated differently. Because naloxone kits may remain available for use across multiple years rather than only in the year dispensed, we represented naloxone as a depreciating stock rather than annual dispensing: each year's stock carried forward 45% of the previous year's stock plus new dispensing, with the remainder lost from circulation through expiration and use. This annual carryover rate corresponds to the monthly kit depreciation of 6.5% estimated by (Zang et al., 2022).

**eTable 3. Prior Distribution Bounds for Calibrated Parameters by County**

| Param. | Description | Allegheny | Philadelphia | Dauphin | Erie | Columbia | Clearfield |
|---|---|---|---|---|---|---|---|
| $\beta_0$ | Intercept: NU to PU | (-9.40, 2.16) | (-15.40, 5.16) | (-10.40, 0.16) | (-10.40, 0.16) | (-10.40, 0.16) | (-8.40, -1.16) |
| $\beta_1$ | Slope: opioid | (0.15, 2.50) | (0.2, 1.00) | (0.30, 0.80) | (0.30, 0.80) | (0.30, 2.80) | (0.30, 1.00) |
| $\beta_2$ | Intercept: OUD to MOUD | (-19.50, -2.19) | (-15.50, -4.19) | (-7.50, 1.19) | (-8.50, 0.19) | (-10.50, 2.19) | (-9.50, 1.19) |
| $\beta_3$ | Slope: buprenorphine | (0.20, 2.20) | (0.10, 1.4) | (0.00, 0.50) | (0.00, 0.70) | (0.00, 0.50) | (0.00, 0.90) |
| $\beta_4$ | Intercept: OUD to OD death | (-12.50, -3.00) | (-9.00, -4.0) | (-8.00, 0.53) | (-9.00, -6.53) | (-8.00, -3.53) | (-7.00, -3.53) |
| $\beta_5$ | Slope: fentanyl | (0.00, 0.40) | (0.00, 1.90) | (0.00, 0.90) | (0.10, 0.90) | (0.00, 0.6) | (0.30, 1.30) |
| $\beta_6$ | Slope: naloxone | (0.00, 0.60) | (0.00, 0.60) | (0.00, 1.50) | (0.05, 0.30) | (0.00, 0.4) | (0.00, 0.90) |

**Note.** Values represent (lower, upper) bounds of uniform prior distributions. NU = nonuser; PU = prescription opioid user; OUD = opioid use disorder.

**eTable 4. Posterior Parameter Estimates from Bayesian Calibration by County**

| County | Parameter | Mean | SD | 95% Uncertainty Interval |
|---|---|---|---|---|
| Allegheny | $\beta_0$ (intercept, NU to PU) | -1.76 | 0.93 | (-3.11, 0.56) |
| | $\beta_1$ (opioid slope) | 1.20 | 0.61 | (0.29, 2.41) |
| | $\beta_2$ (intercept, OUD to MOUD) | -9.03 | 5.22 | (-18.95, -2.90) |
| | $\beta_3$ (buprenorphine slope) | 1.26 | 0.55 | (0.29, 2.14) |
| | $\beta_4$ (intercept, OUD to OD death) | -7.44 | 0.10 | (-7.62, -7.25) |
| | $\beta_5$ (fentanyl slope) | 0.05 | 0.04 | (0.00, 0.15) |
| | $\beta_6$ (naloxone slope) | 0.32 | 0.07 | (0.19, 0.45) |
| Philadelphia | $\beta_0$ (intercept, NU to PU) | -3.19 | 0.22 | (-3.61, -2.77) |
| | $\beta_1$ (opioid slope) | 0.34 | 0.12 | (0.20, 0.66) |
| | $\beta_2$ (intercept, OUD to MOUD) | -11.31 | 2.59 | (-15.35, -6.33) |
| | $\beta_3$ (buprenorphine slope) | 0.75 | 0.36 | (0.16, 1.36) |
| | $\beta_4$ (intercept, OUD to OD death) | -6.85 | 0.07 | (-7.00, -6.73) |
| | $\beta_5$ (fentanyl slope) | 0.03 | 0.03 | (0.00, 0.10) |
| | $\beta_6$ (naloxone slope) | 0.05 | 0.04 | (0.00, 0.13) |
| Dauphin | $\beta_0$ (intercept, NU to PU) | -7.21 | 1.75 | (-10.24, -4.29) |
| | $\beta_1$ (opioid slope) | 0.54 | 0.14 | (0.31, 0.78) |
| | $\beta_2$ (intercept, OUD to MOUD) | -5.58 | 1.22 | (-7.43, -3.15) |
| | $\beta_3$ (buprenorphine slope) | 0.25 | 0.14 | (0.01, 0.49) |
| | $\beta_4$ (intercept, OUD to OD death) | -7.50 | 0.18 | (-7.79, -7.04) |
| | $\beta_5$ (fentanyl slope) | 0.21 | 0.12 | (0.02, 0.47) |
| | $\beta_6$ (naloxone slope) | 0.14 | 0.08 | (0.01, 0.31) |
| Erie | $\beta_0$ (intercept, NU to PU) | -2.34 | 1.06 | (-4.13, -0.07) |
| | $\beta_1$ (opioid slope) | 0.55 | 0.14 | (0.31, 0.79) |
| | $\beta_2$ (intercept, OUD to MOUD) | -3.99 | 2.25 | (-8.21, -0.69) |
| | $\beta_3$ (buprenorphine slope) | 0.35 | 0.19 | (0.02, 0.68) |
| | $\beta_4$ (intercept, OUD to OD death) | -7.39 | 0.40 | (-8.05, -6.60) |
| | $\beta_5$ (fentanyl slope) | 0.28 | 0.11 | (0.11, 0.52) |
| | $\beta_6$ (naloxone slope) | 0.17 | 0.07 | (0.06, 0.29) |
| Columbia | $\beta_0$ (intercept, NU to PU) | -5.04 | 2.69 | (-10.07, -0.12) |
| | $\beta_1$ (opioid slope) | 1.53 | 0.71 | (0.36, 2.71) |
| | $\beta_2$ (intercept, OUD to MOUD) | -4.53 | 3.65 | (-10.20, 1.78) |
| | $\beta_3$ (buprenorphine slope) | 0.25 | 0.14 | (0.02, 0.49) |
| | $\beta_4$ (intercept, OUD to OD death) | -7.13 | 0.62 | (-7.94, -5.68) |

| County | Parameter | Mean | SD | 95% Uncertainty Interval |
|---|---|---|---|---|
| | $\beta_5$ (fentanyl slope) | 0.34 | 0.16 | (0.03, 0.59) |
| | $\beta_6$ (naloxone slope) | 0.22 | 0.10 | (0.03, 0.39) |
| Clearfield | $\beta_0$ (intercept, NU to PU) | -4.63 | 1.62 | (-8.08, -2.31) |
| | $\beta_1$ (opioid slope) | 0.65 | 0.20 | (0.32, 0.98) |
| | $\beta_2$ (intercept, OUD to MOUD) | -2.18 | 0.77 | (-3.91, -1.24) |
| | $\beta_3$ (buprenorphine slope) | 0.28 | 0.22 | (0.01, 0.81) |
| | $\beta_4$ (intercept, OUD to OD death) | -6.67 | 0.27 | (-6.99, -5.99) |
| | $\beta_5$ (fentanyl slope) | 0.93 | 0.23 | (0.44, 1.28) |
| | $\beta_6$ (naloxone slope) | 0.25 | 0.19 | (0.01, 0.71) |

**Note.** Estimates from IMIS with 250 resampled parameter sets. 95% uncertainty intervals are 2.5th and 97.5th percentiles of the posterior distribution.

## eFigure 1. Model Calibration and Validation Results by County

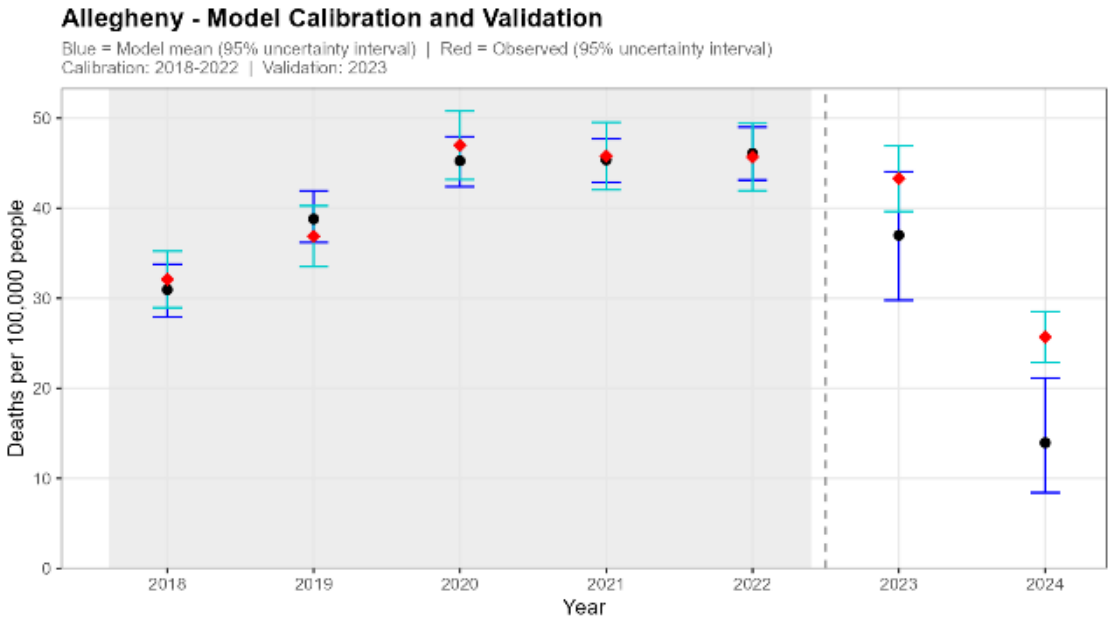

*(a) Allegheny: Model predictions vs. calibration targets*

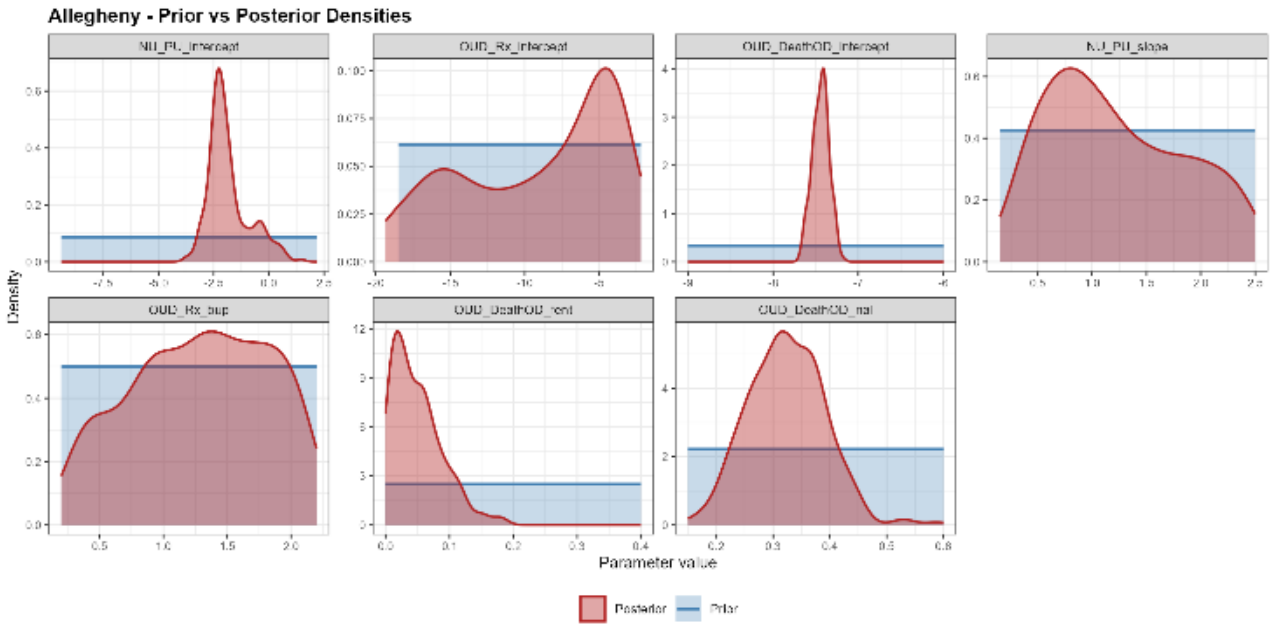

*(b) Allegheny: Prior and posterior distributions of calibrated parameters*

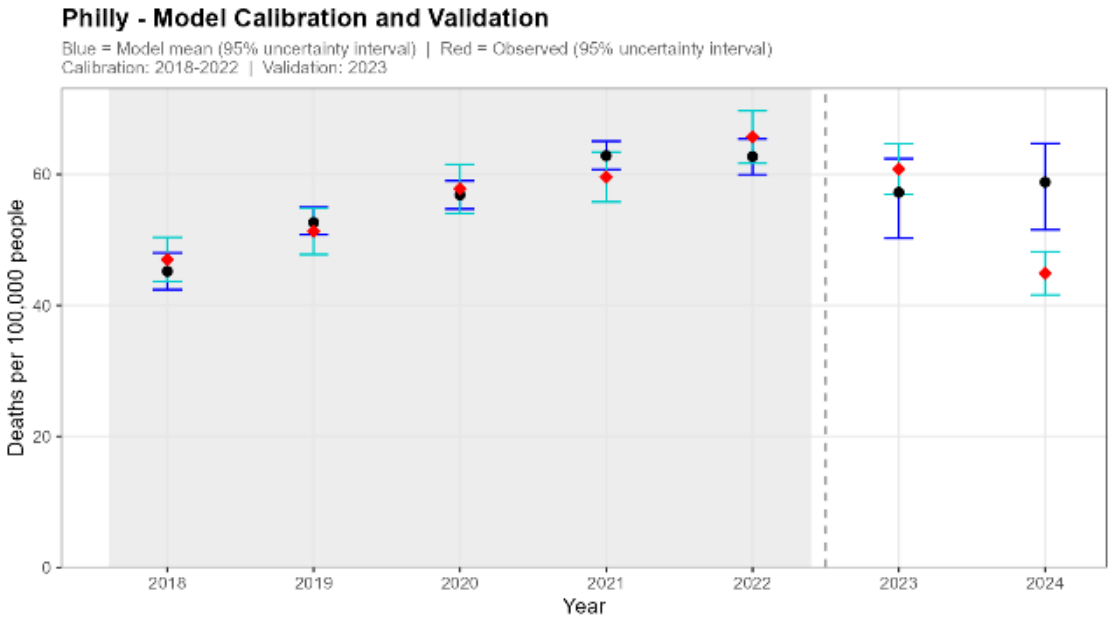

*(a) Philadelphia: Model predictions vs. calibration targets*

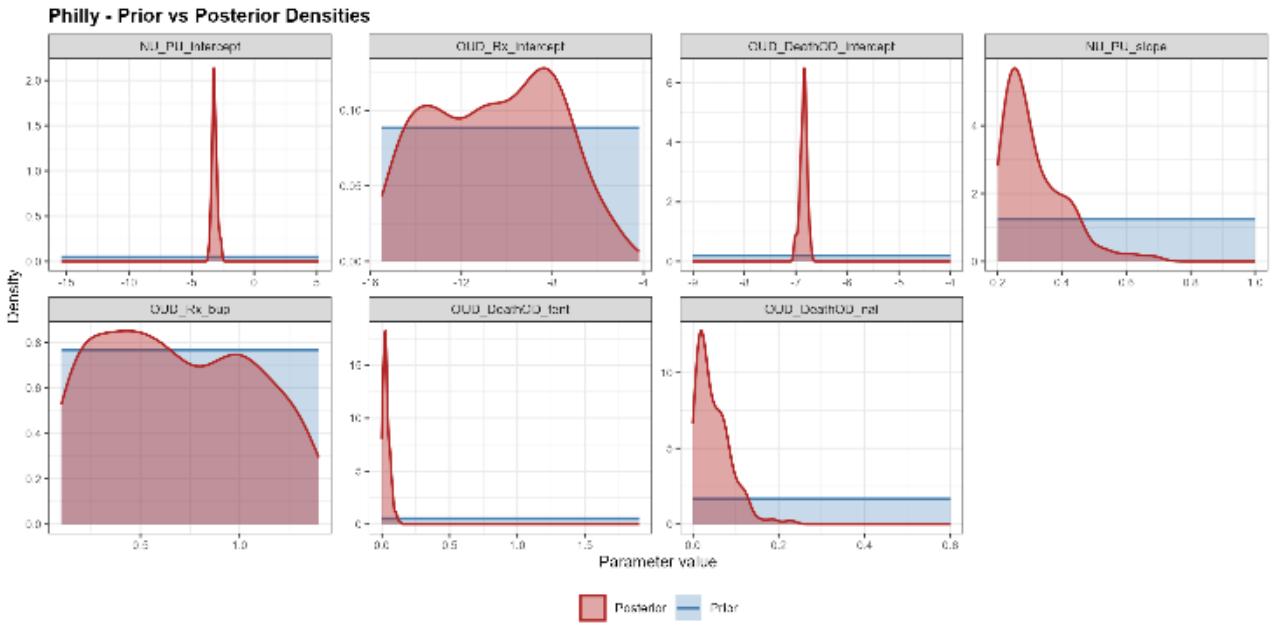

*(b) Philadelphia: Prior and posterior distributions of calibrated parameters*

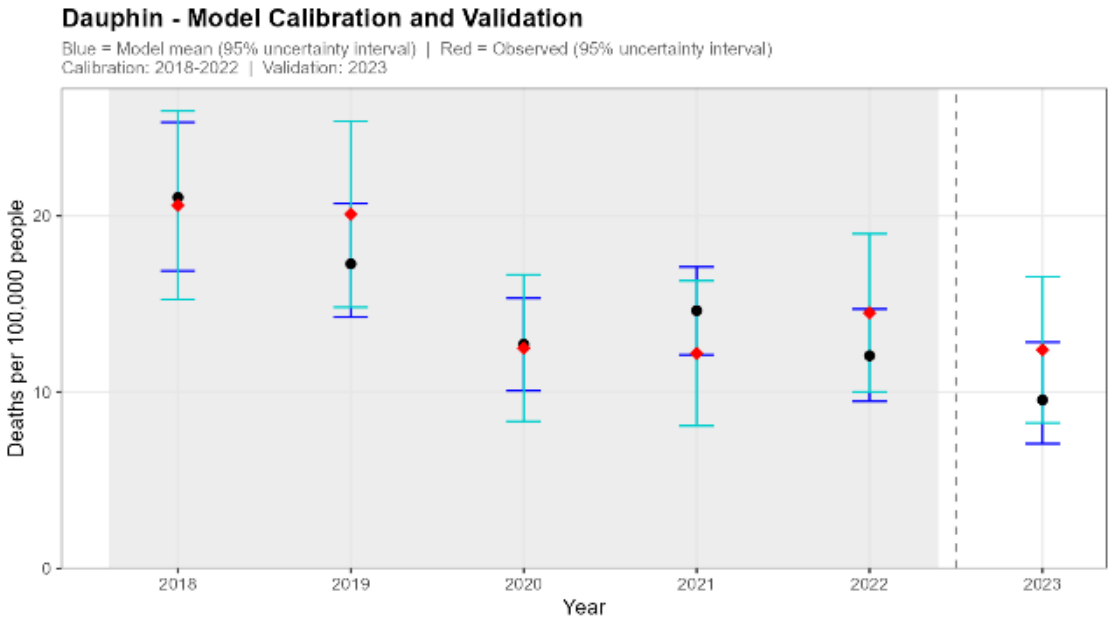

*(a) Dauphin: Model predictions vs. calibration targets*

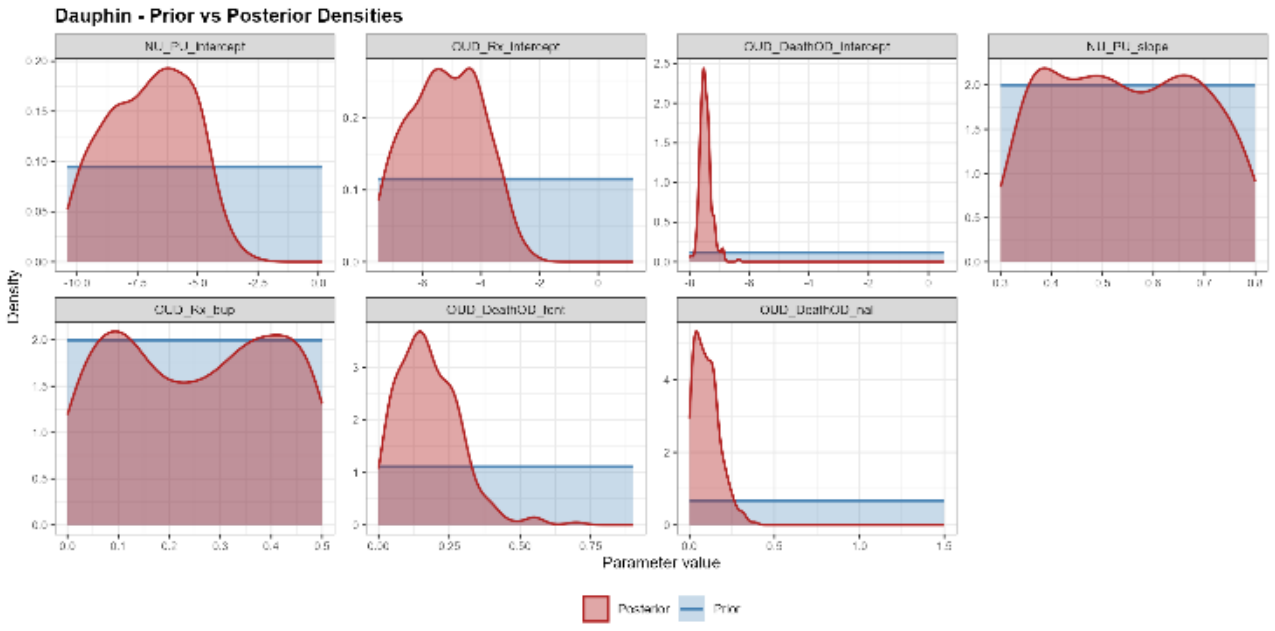

*(b) Dauphin: Prior and posterior distributions of calibrated parameters*

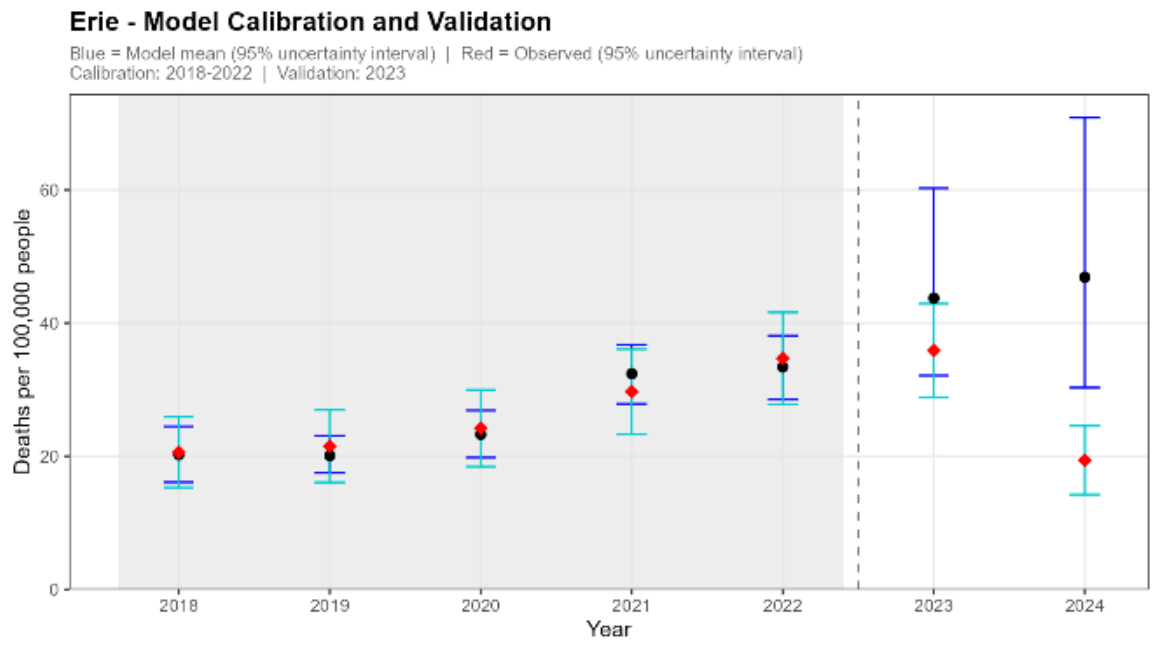

(a) *Erie: Model predictions vs. calibration targets*

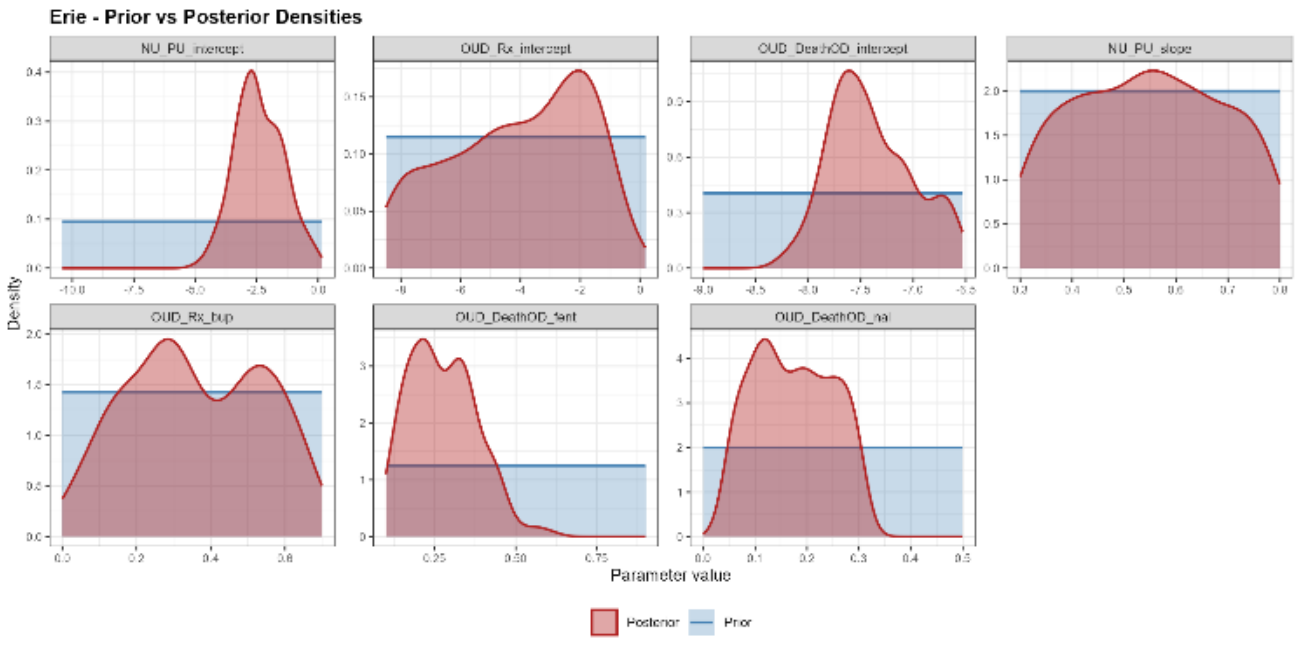

(b *Erie: Prior and posterior distributions of calibrated parameters*

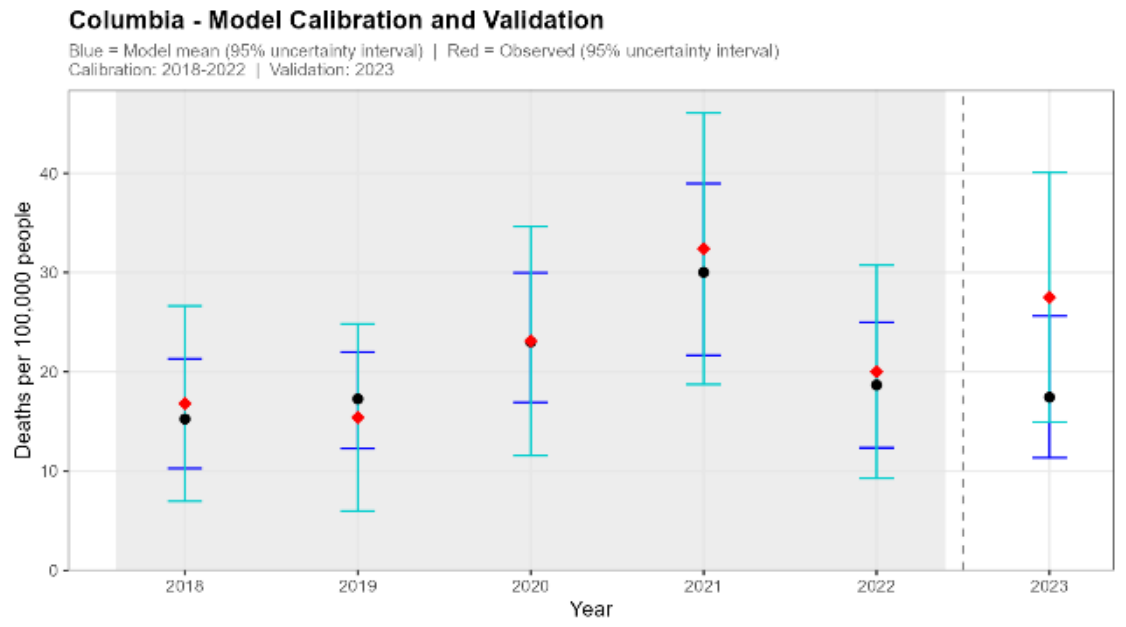

(a) Columbia: Model predictions vs. calibration targets

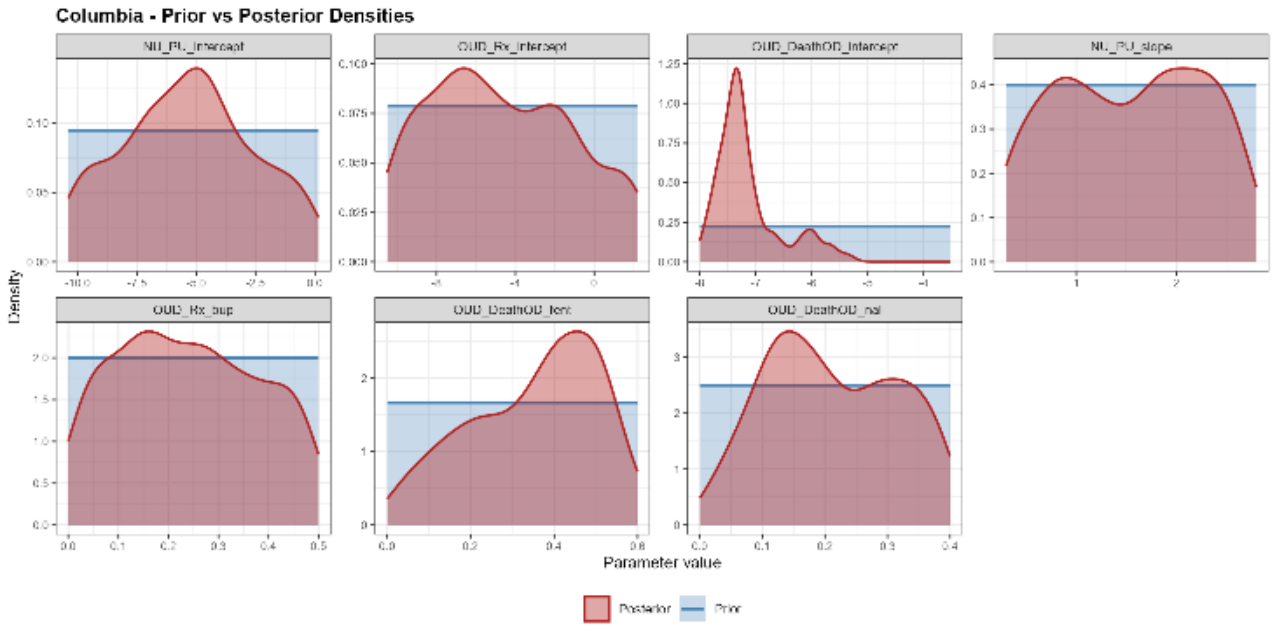

(b) Columbia: Prior and posterior distributions of calibrated parameters

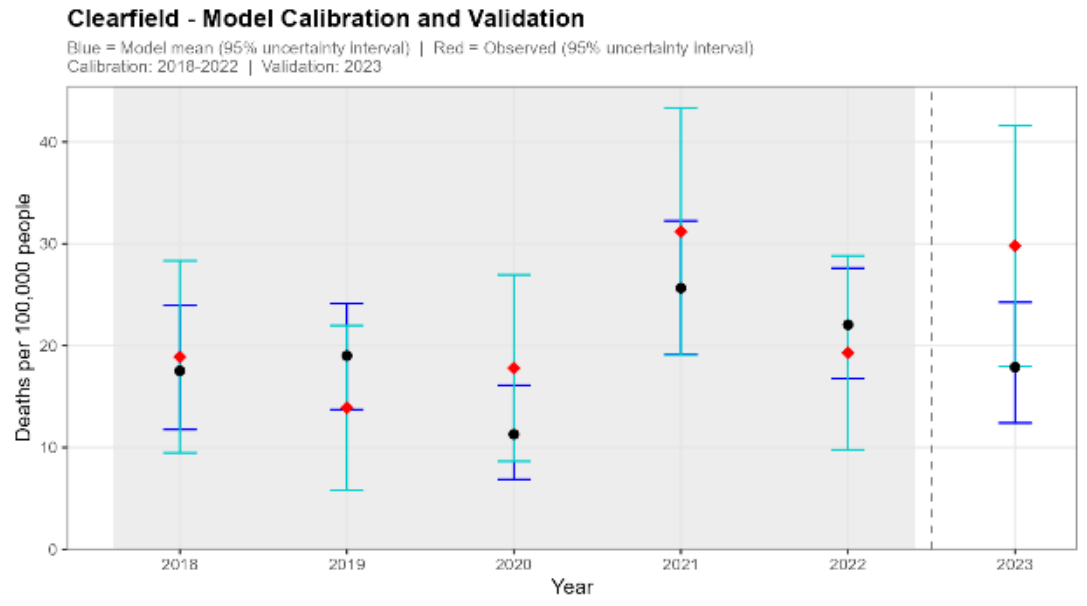

*(a) Clearfield: Model predictions vs. calibration targets*

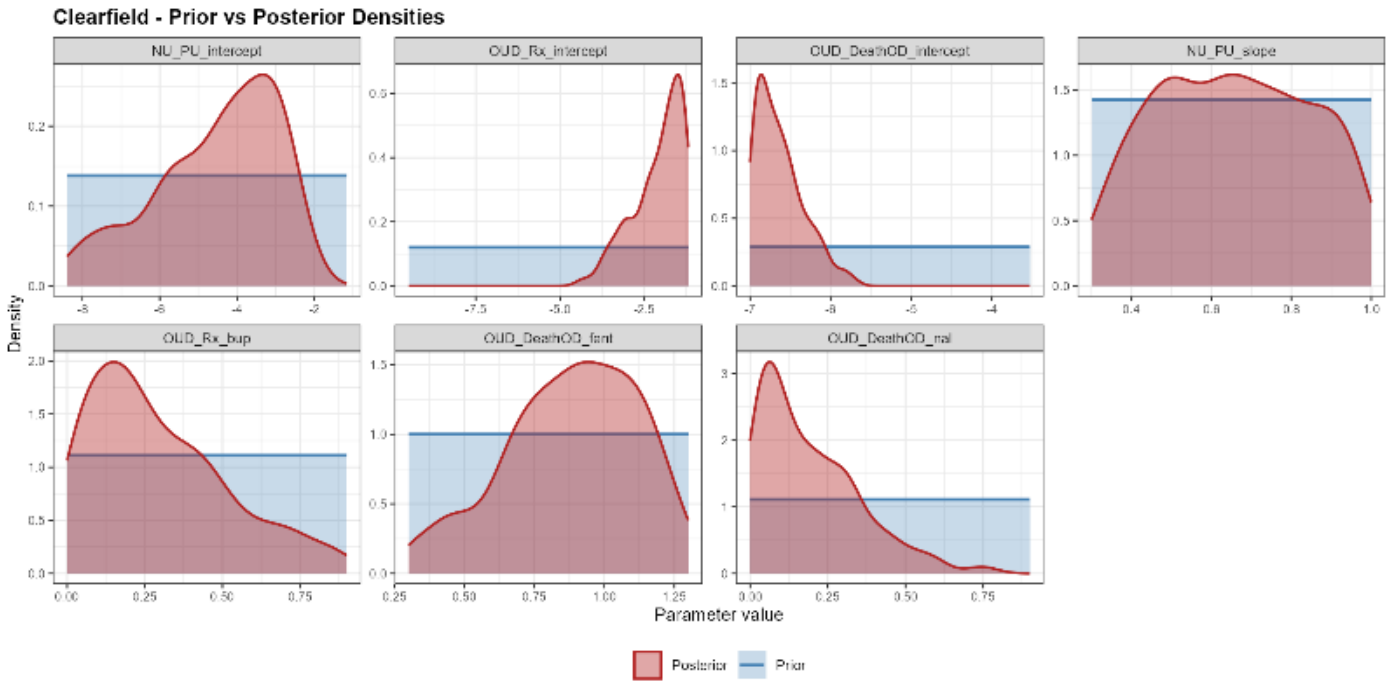

*(b) Clearfield: Prior and posterior distributions of calibrated parameters*

*eFigure 1. Each panel shows one county. Left: Observed (points) versus model-predicted (line, 95% uncertainty interval) overdose death rates per 100,000, 2018-2024. Model calibrated to 2018-2022; validated against 2023-2024 (out-of-sample) except for Columbia and Clearfield (validation against 2023). Right: Prior (blue) and posterior (red) distributions for the seven calibrated parameters.*

**eFigure 2. County-Specific Naloxone and Buprenorphine Dispensing Rates, 2018–2029**

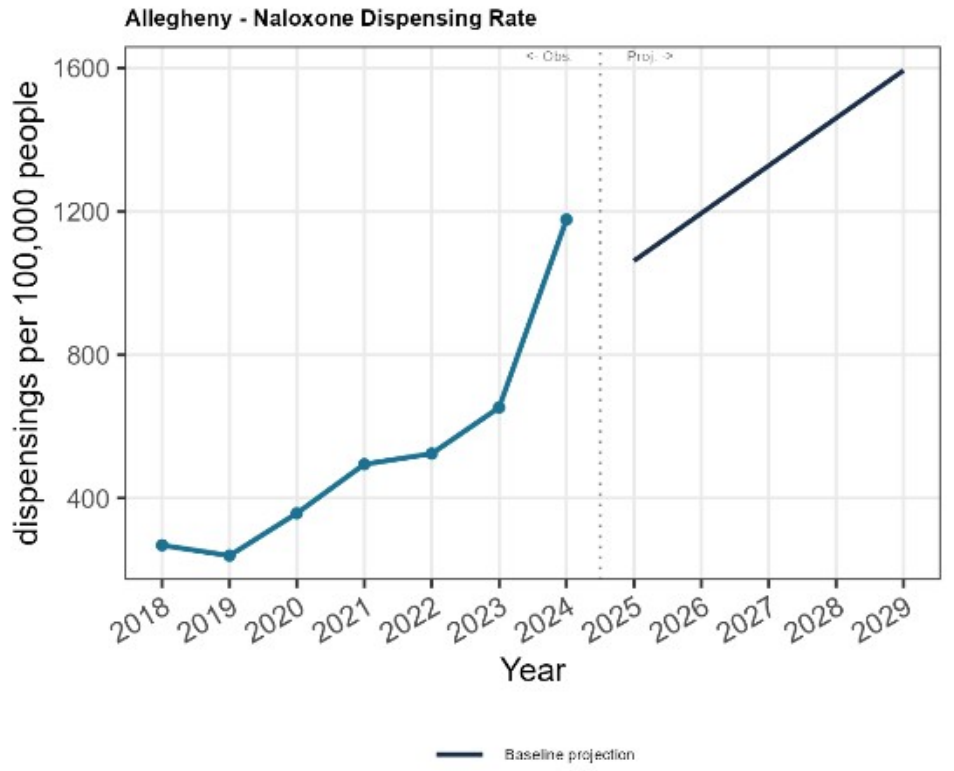

*(a) Allegheny: Naloxone dispensing rate*

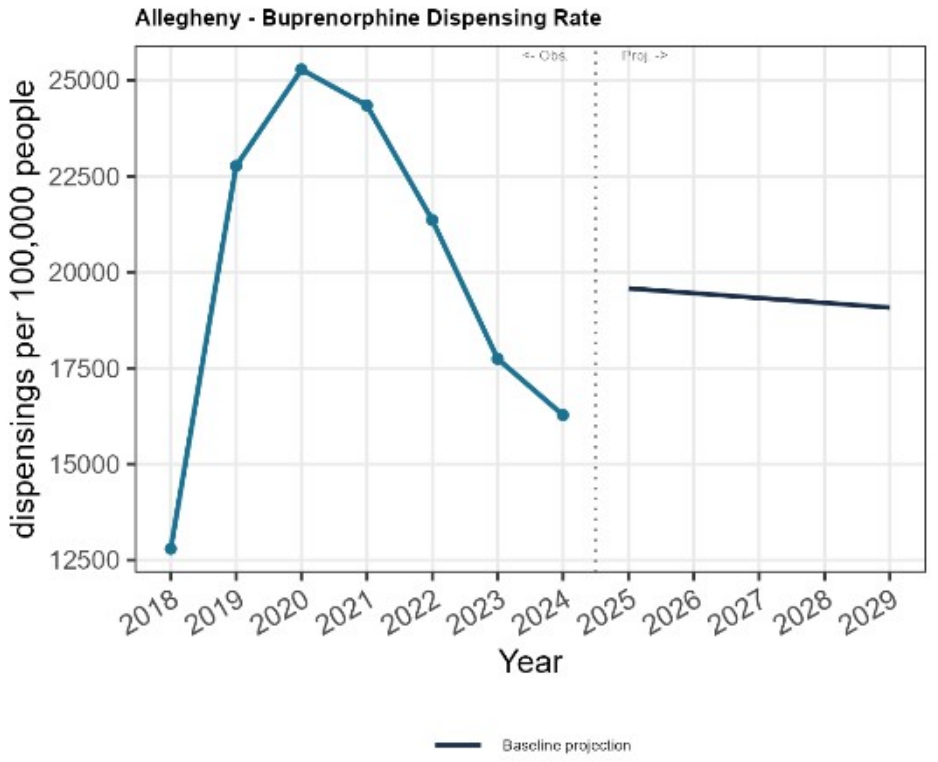

*(b) Allegheny: Buprenorphine dispensing rate*

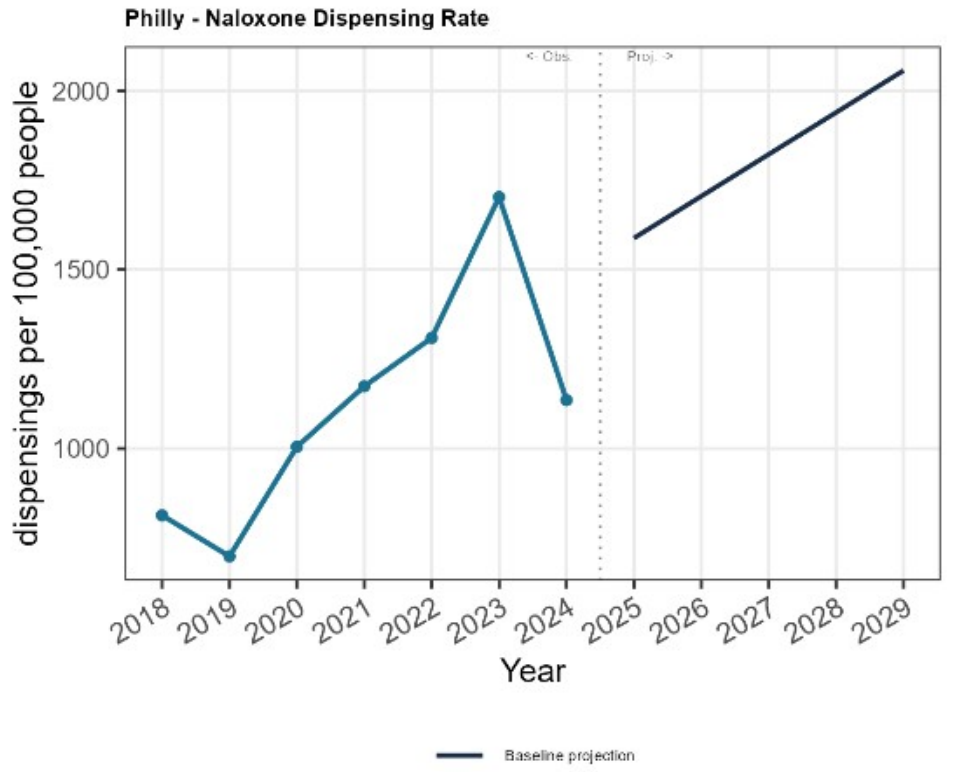

*(a) Philadelphia: Naloxone dispensing rate*

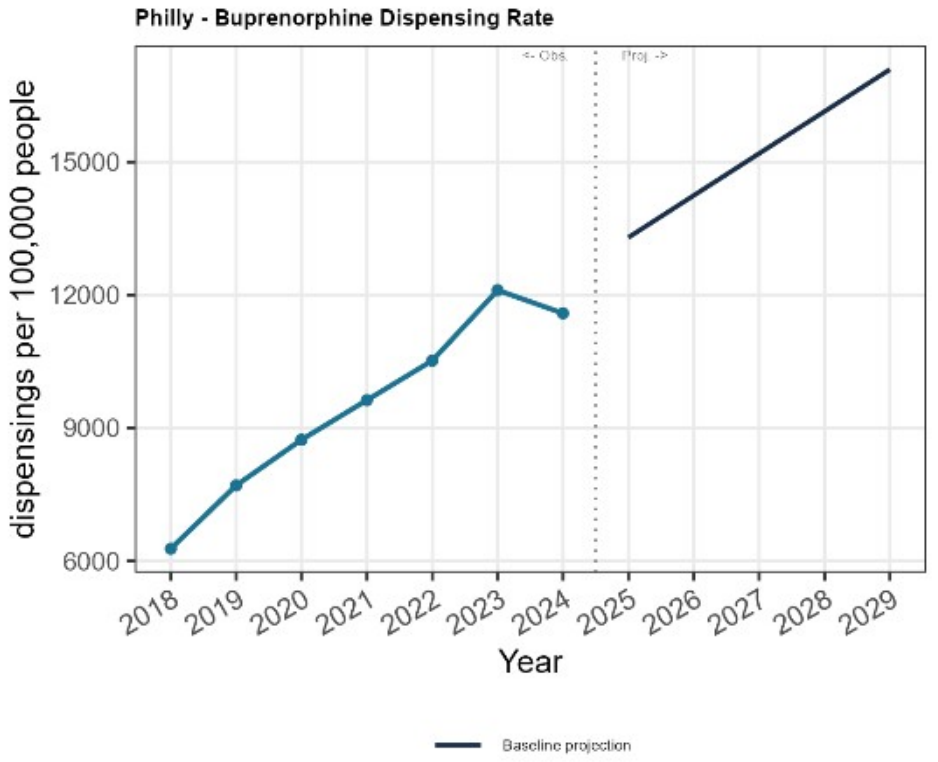

*(b) Philadelphia: Buprenorphine dispensing rate*

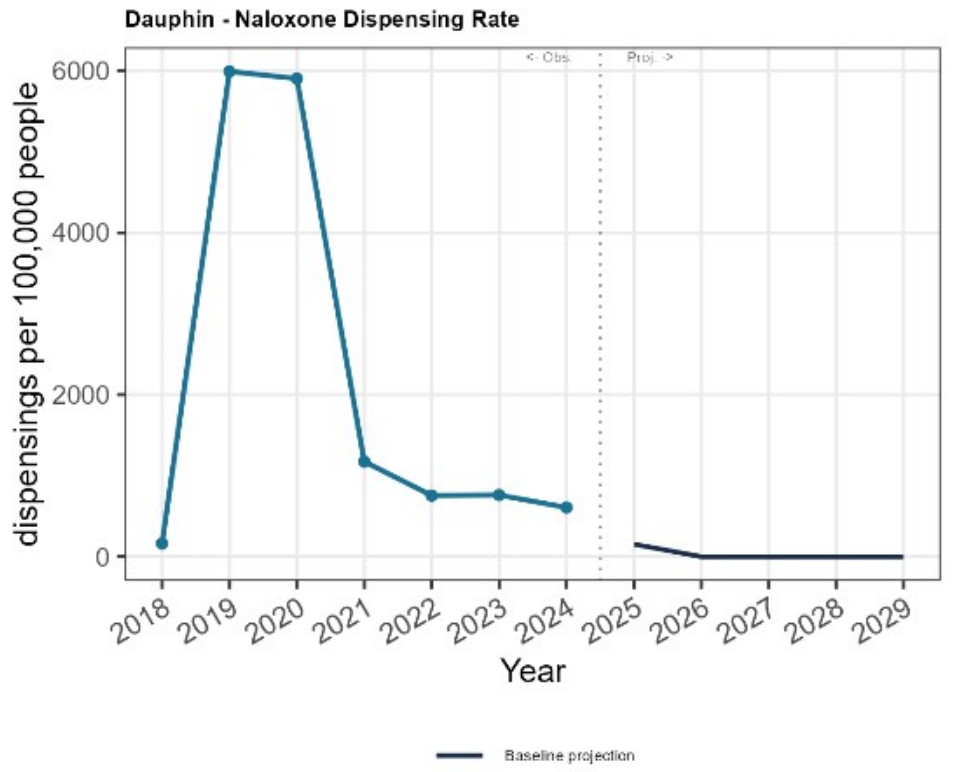

*(a) Dauphin: Naloxone dispensing rate*

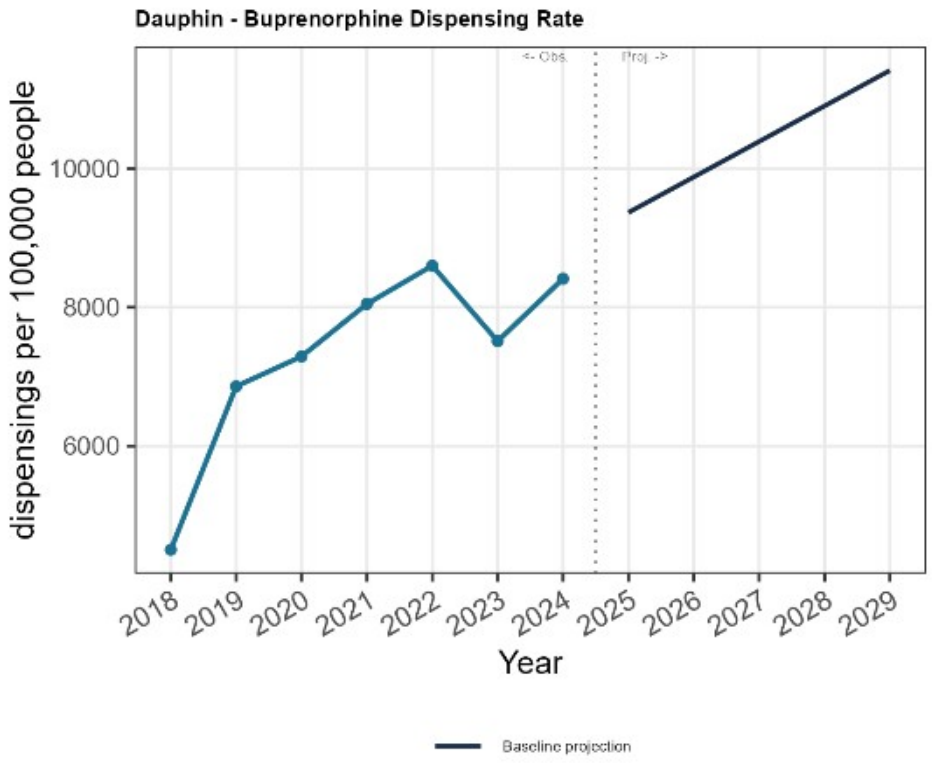

*(b) Dauphin: Buprenorphine dispensing rate*

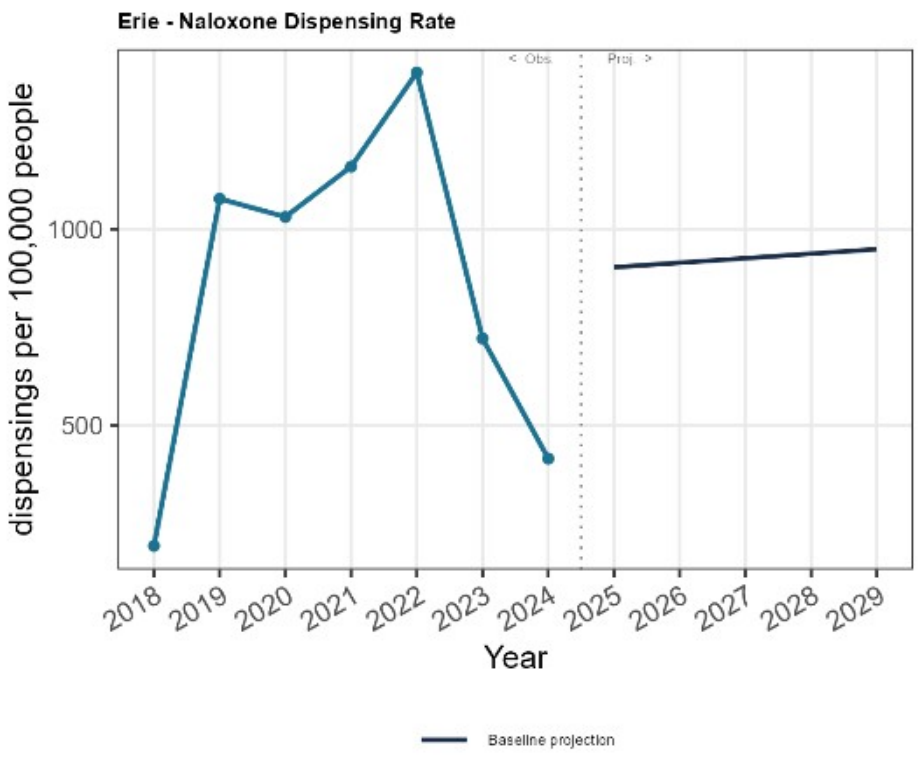

*(a) Erie: Naloxone dispensing rate*

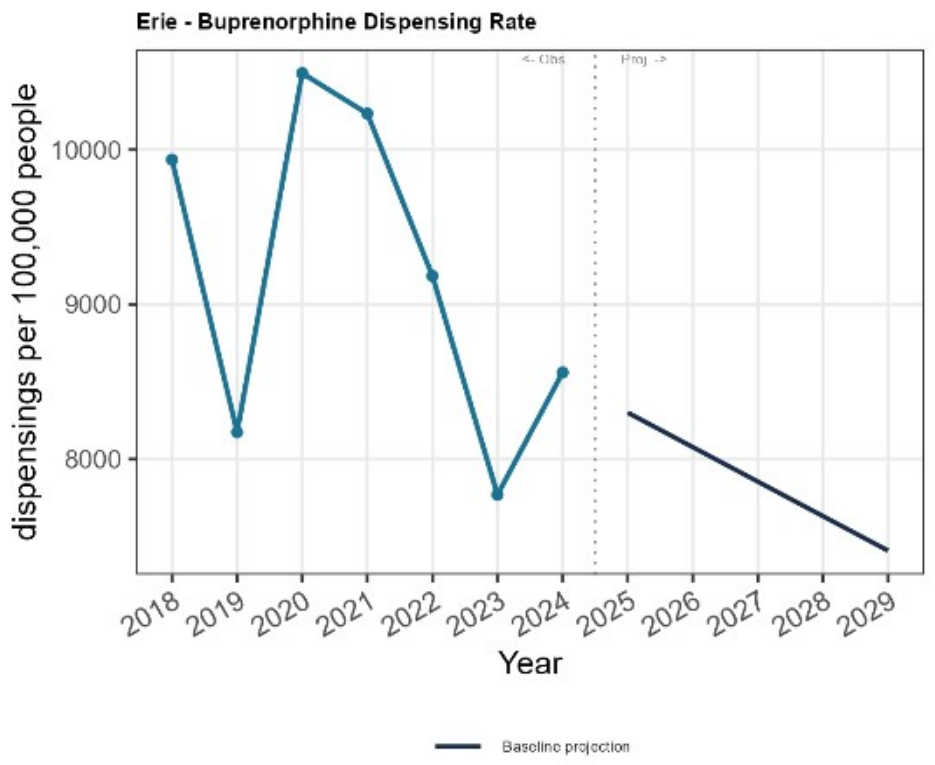

*(b) Erie: Buprenorphine dispensing rate*

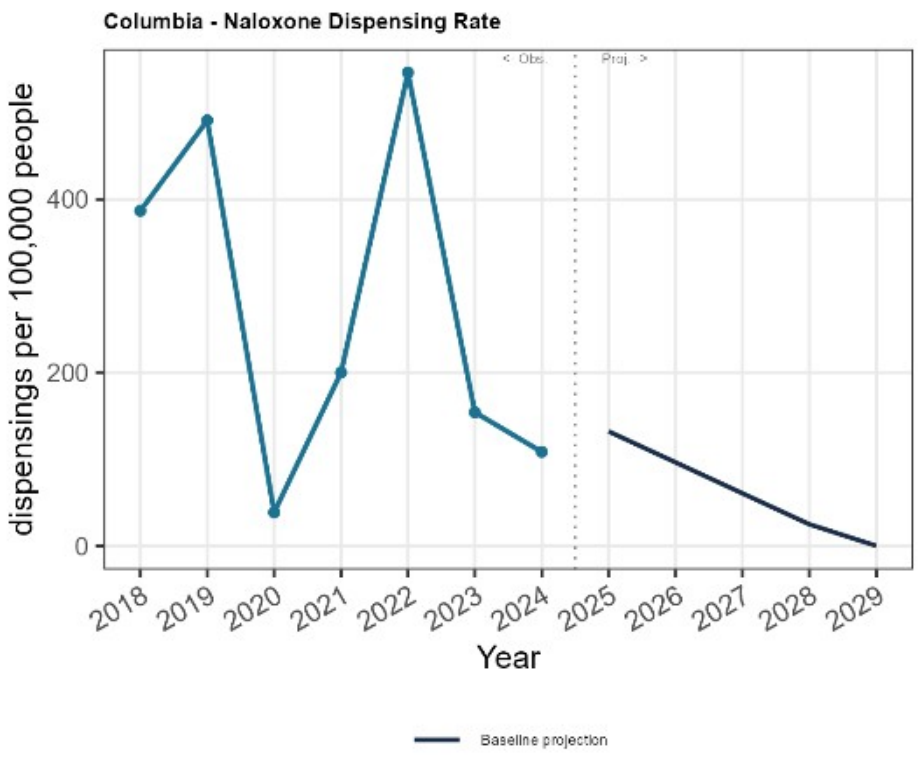

*(a) Columbia: Naloxone dispensing rate*

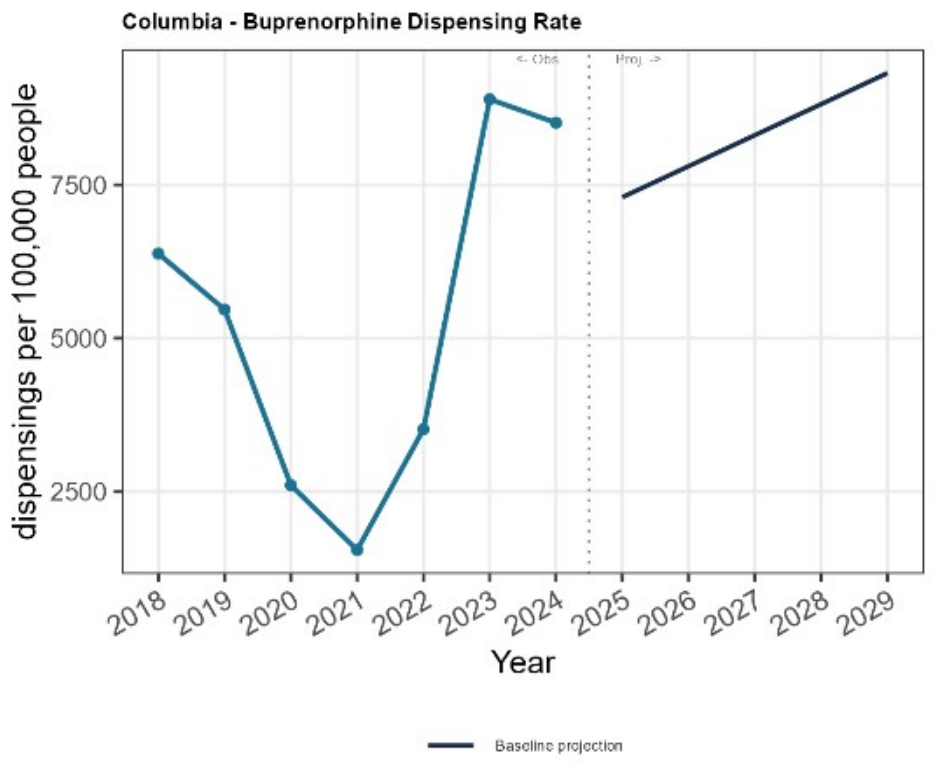

*(b) Columbia: Buprenorphine dispensing rate*

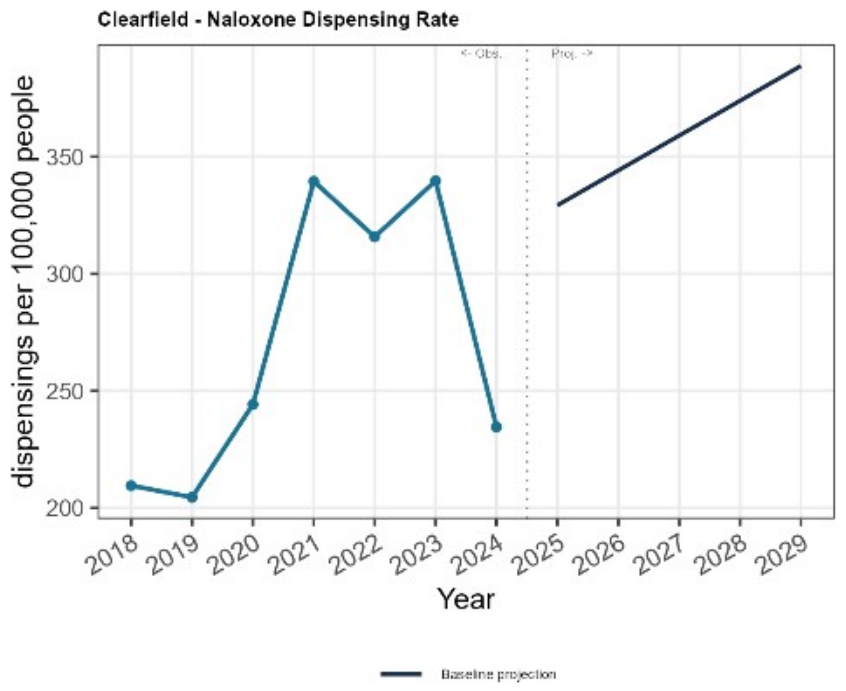

*(a) Clearfield: Naloxone dispensing rate*

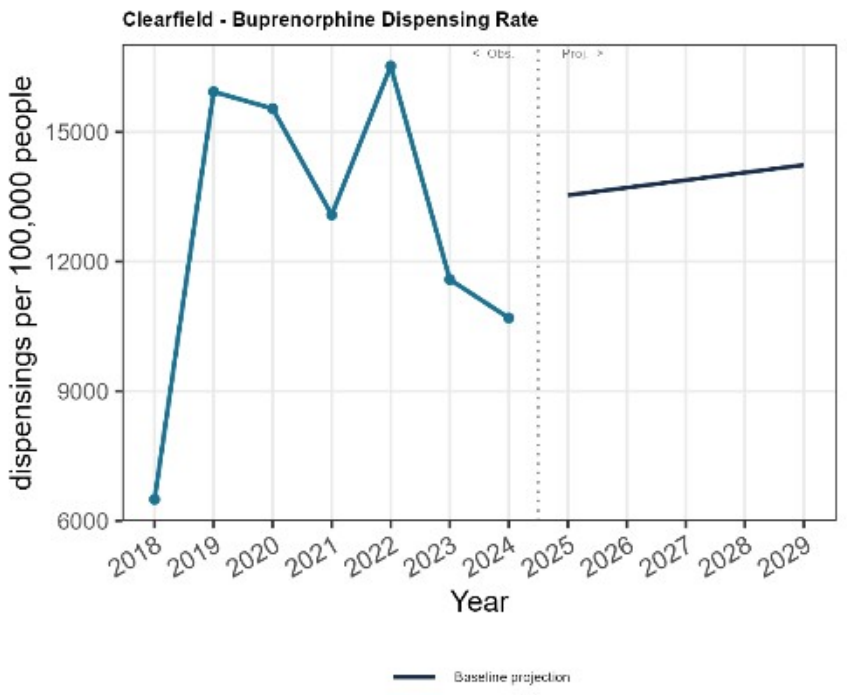

*(b) Clearfield: Buprenorphine dispensing rate*

*eFigure 2. Each panel shows the annual dispensing rate (per 100,000 population) of one intervention covariate in one county over the 2018–2024 period and projection for 2025-2029 period. Projected values were obtained by fitting a linear trend to the observed 2018–2024 dispensing rates and extrapolating forward. Left column: naloxone dispensing rates; right column: buprenorphine dispensing rates. Rows correspond to Allegheny, Philadelphia, Dauphin, Erie, Columbia, and Clearfield. Source: IQVIA dispensing data.*

## eAppendix 3: Allegheny Complementary Calibration

In the primary analysis, Allegheny was calibrated to overdose mortality over 2018-2022, consistent with the other counties. This calibration under-predicted observed 2023-2024 mortality, producing a low projected baseline. As a robustness check, we recalibrated Allegheny [including 2018-2024 as calibration targets]. eFigure 3 shows the resulting calibration and fit. Under this complementary calibration, the projected 2029 baseline overdose death rate was 5 (95% UI: 2.9-8.3) per 100,000 people.

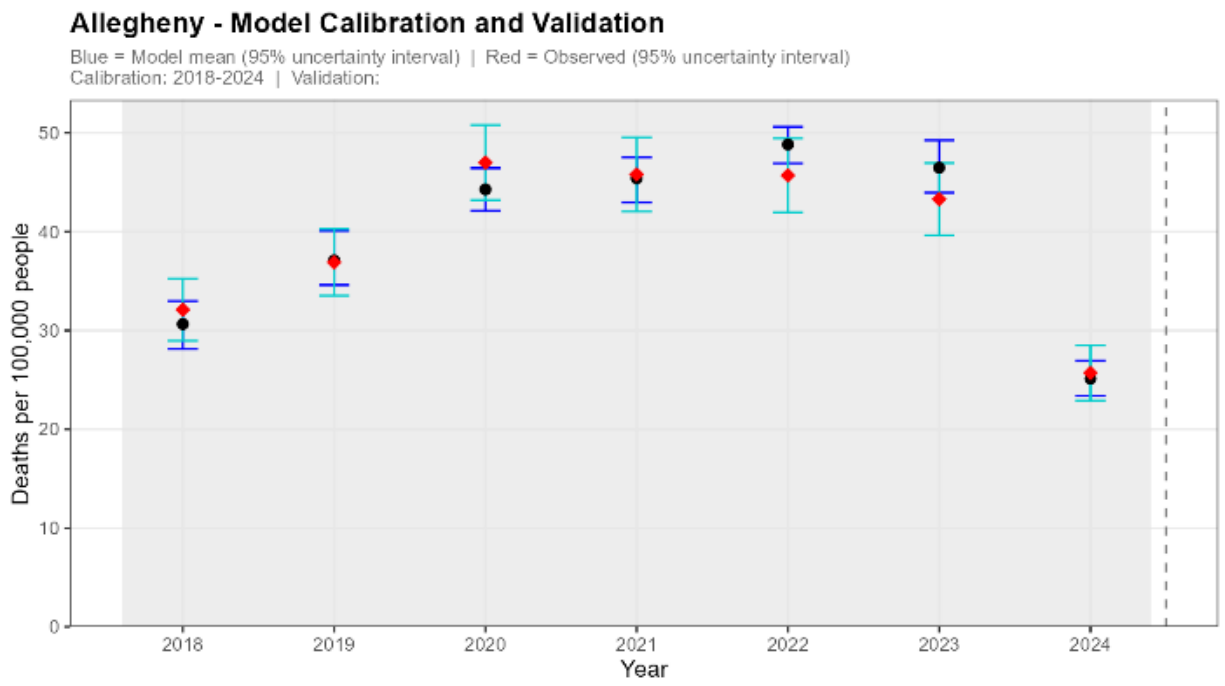

*(a) Allegheny: Model predictions vs. calibration targets*

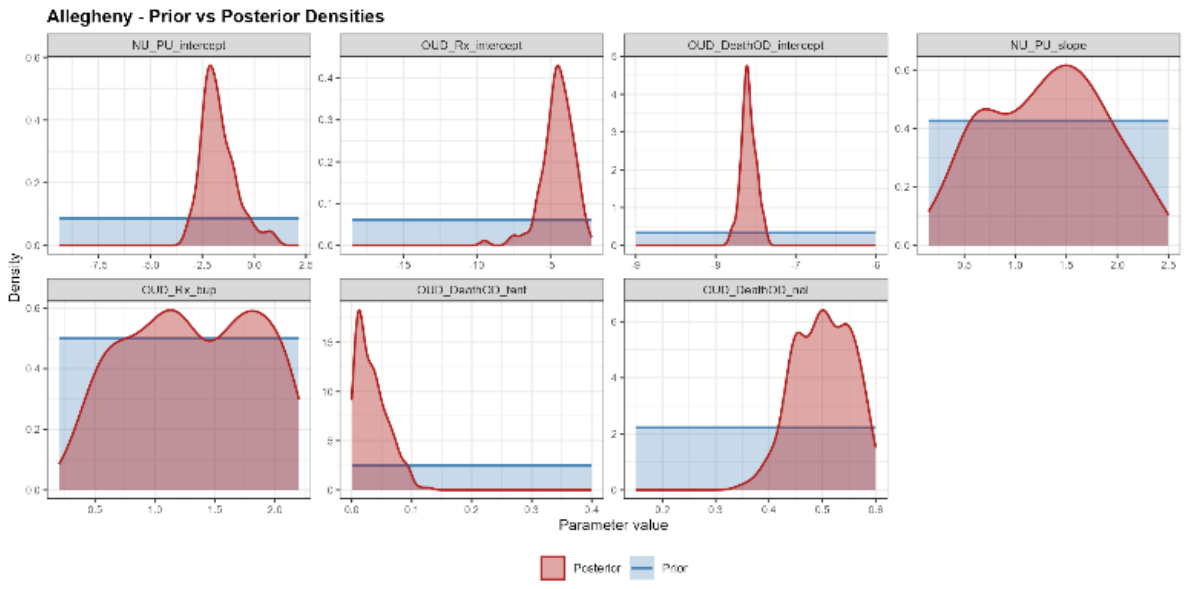

*(b) Allegheny: Prior and posterior distributions of calibrated parameters*

*eFigure 3. Allegheny complementary calibration. Left: Observed (points) versus model-predicted (line, 95% uncertainty interval) overdose death rates per 100,000, 2018-2024. Model calibrated to 2018-2024 (all available years; no out-of-sample validation year). Right: Prior (blue) and posterior (red) distributions for the seven calibrated parameters.*

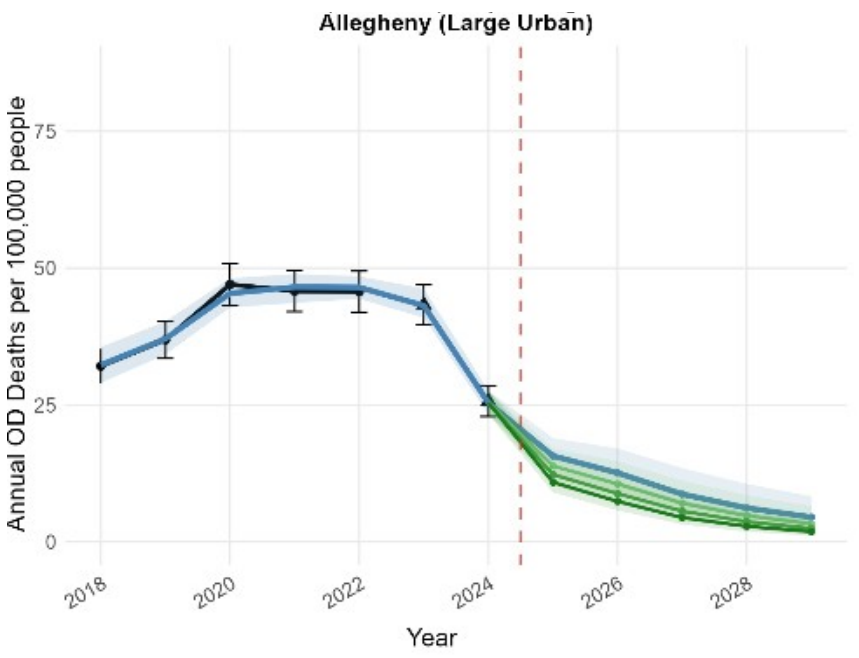

*(a) Allegheny, 2018-2024 calibration: projected overdose deaths under naloxone scale-up (+10/20/30%), 2025-2029.*

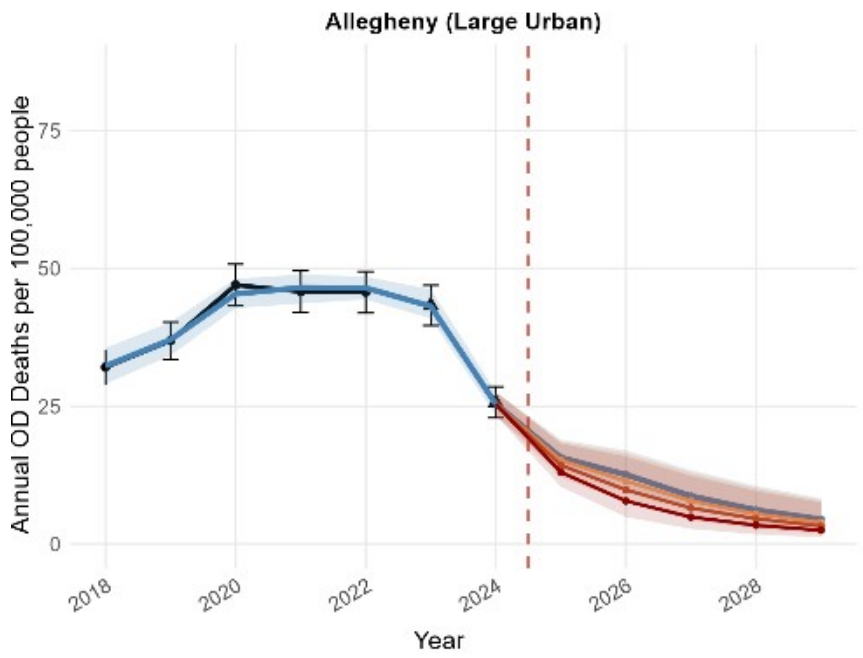

*(b) Allegheny, 2018-2024 calibration: projected overdose deaths under buprenorphine scale-up (+10/20/30%), 2025-2029.*

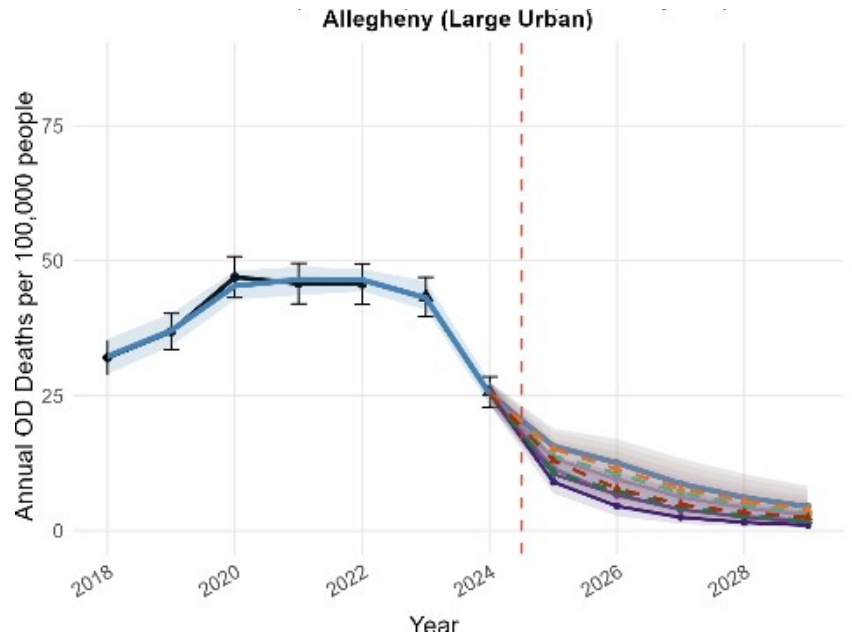

*(c) Allegheny, 2018-2024 calibration: projected overdose deaths under combined naloxone + buprenorphine scale-up, 2025-2029.*

eFigure 4. Projected overdose death trajectories for Allegheny County under the complementary 2018-2024 calibration. Observed rates (2018-2024), model baseline, and intervention scenarios (naloxone, buprenorphine, and combined; +10%/+20%/+30% above projected baseline dispensing), 2025-2029. Shaded bands are 95% posterior intervals. Compare with the primary-calibration projection in Figures 2-4.